\documentclass[useAMS,usenatbib,usegraphicx]{mn2e}
\usepackage{multirow}
\usepackage{url}
\usepackage{rotating}

\title[Swift/XRT observations of unidentified INTEGRAL/IBIS sources] 
{\emph{Swift}/XRT observations of unidentified \emph{INTEGRAL}/IBIS sources}

\author[R. Landi et al.]
{R. Landi,$^1$\thanks{E-mail address: \texttt{landi@iasfbo.inaf.it}}, L.~Bassani$^1$, A.~Malizia$^1$,
J.B. Stephen$^1$, A.~Bazzano$^2$, M.~Fiocchi$^2$, \newauthor A.J.~Bird$^3$ \\
$^1$ INAF/IASF-Bologna, Via P. Gobetti 101, I-40129 Bologna, Italy \\
$^2$ INAF/IASF-Roma, Via Fosso del Cavaliere 100, I-00133, Roma, Italy \\
$^3$ School of Physics and Astronomy, University of Southampton,
        SO17 1BJ, Southampton, UK\\
}

\date{Accepted .... Received ...; in original form ...}
\pagerange{\pageref{firstpage}--\pageref{lastpage}} \pubyear{2009}

\begin{document}
\label{firstpage}
\maketitle 

\begin{abstract}
The most recent IBIS/ISGRI survey, i.e. the fourth one, lists 723 hard X-ray sources many still 
unidentified, i.e. lacking an X-ray counterpart or simply not studied at lower energies, i.e. below 
10 keV. In order to overcome this lack of X-ray information, we cross-correlated the list of IBIS 
sources included in the fourth IBIS catalogue with the \emph{Swift}/XRT data archive, finding a sample 
of 20 objects, not yet reported in the literature, for which XRT data could help in the search for the 
X-ray and hence optical counterpart and/or, for the first time, in the study of the source spectral and 
variability properties below 10 keV.
Sixteen of these objects are new \emph{INTEGRAL} detections, while four were already listed in the third 
survey but not yet observed in X-rays.
Four objects (IGR J00465--4005, LEDA 96373, IGR J1248.2--5828 and IGR J13107--5626) are confirmed or likely
absorbed active galaxies, while two (IGR J14080--3023 and 1RXS J213944.3+595016) are unabsorbed AGN.
We also find three peculiar extragalactic objetcs, NGC 4728 being a Narrow Line Seyfert galaxy,
MCG+04--26--006 a type 2 LINER and PKS 1143--693 probably a QSO; furthermore, our results indicate that  
IGR J08262+4051 and IGR J22234--4116 are candidate AGN, which
require further optical spectroscopic follow-up observations to be fully classified.
Only in the case of 1RXS J080114.6--462324 we are confident that the source is a Galactic object.
For IGR J10447--6027, IGR J12123--5802 and IGR J20569+4940 we pinpoint one X-ray counterpart, 
although its nature could not be assessed despite spectral and sometimes variability
information being obtained. Clearly, we need to perform optical follow-up observations in order to firmly 
assess their nature.
There are five objects for which we find no obvious X-ray counterpart (IGR J07506--1547 and IGR 
J17008--6425) or even no detection (IGR J17331--2406, IGR J18134--1636 and IGR J18175--1530); apart from 
IGR J18134--1636, all these sources are found to be variable in the IBIS energy band, therefore it is 
difficult to catch them even in X-rays.
\end{abstract}


\section{Introduction} 
In recent years, our knowledge of the hard X-ray sky ($>$10 keV) has improved 
significantly thanks to the results obtained by IBIS (Ubertini et al. 2003) on board \emph{INTEGRAL} 
(Winkler et al. 2003) and BAT (Barthelmy et al. 2005) on board \emph{Swift} (Gehrels et al. 2004). Both 
telescopes operate in similar wave bands (around 20--200 keV) with a limiting sensitivity of about a mCrab 
and a point source location accuracy of the order of a few arcminutes. These instruments continue to 
perform a survey of the high energy sky, thus providing the best yet sample of objects selected in the soft 
gamma-ray band. A significant number of the sources detected by these two satellites are still 
unidentified/unclassified and/or have no information in the 2--10 keV energy range. Recently, the fourth 
\emph{INTEGRAL}/IBIS survey (Bird et al. 2009) containing 723 hard X-ray (17--100 keV) emitters has been 
compiled. For many sources in this new IBIS catalogue a refined localisation, which is possible by 
exploiting the capability of current focusing X-ray telescopes, is necessary in order to pinpoint and 
classify their optical counterpart. Furthermore, information in the X-ray band are often lacking so that it 
is not always possible to characterise these sources in terms of spectral shape, flux, absorption 
properties and variability. To this aim, we have cross-correlated the list of IBIS sources included in 
the fourth IBIS catalogue with the archive of all \emph{Swift}/XRT pointings, finding a sample of 20 
objects for which low energy data can help in identifying the X-ray and hence the optical counterpart 
and/or in providing X-ray spectral and variability information. We emphasise that most of the associations
and hence types listed in the fourth IBIS survey, for these 20 sources, stem from this work.

The paper is structured as follows: in Sect. 2 we briefly present the method 
adopted for the XRT data reduction and the criteria assumed for the spectral analysis. Sect. 3 is 
devoted to the discussion of the results for each individual sources. Conclusion are drawn in Sect. 4.

\section{Data reduction and analysis}

For all sources in the sample, we use X-ray data acquired with the X-ray Telescope (XRT, 
0.3--10 keV, Burrows et al. 2005) on board the \emph{Swift} satellite.
The XRT data reduction was performed using the XRTDAS standard data pipeline package ({\sc xrtpipeline} 
v. 0.12.2), in order to produce screened event files. All data were extracted only in the Photon Counting 
(PC) mode (Hill et al. 2004), adopting the standard grade filtering (0--12 for PC) according to the XRT 
nomenclature. The log of all XRT observations presented in this paper is given in Table~\ref{tab1}. For 
each measurement, we report the observation code (ID), the observation date and the exposure time. 

For each observation we analysed, with {\sc XIMAGE} v. 4.4.1, the 0.3--10 keV image to search for sources 
detected (at a confidence level $>$ 3$\sigma$) both within the 90\% and 99\% IBIS error circles.
Then, we estimated the X-ray positions using the task {\sc xrtcentroid v.0.2.9} by taking into account 
the longest pointing for those sources with more than one observation. XRT images are shown in Figure 1 to 
20 with overlapping IBIS error circles and, when available, also BAT positional uncertainties. 
Table~\ref{tab2} lists all 20 IBIS sources analysed here together with their position and relative uncertainty 
as listed in  Bird et al. (2009) as well as their location with respect to the Galactic plane. For each of 
these gamma-ray emitters, we then report the position and relative uncertainties (at 90\% confidence level) of 
all sources detected by XRT within the 90\% and 99\% IBIS error circles as well as the count rate in the 
0.3--10 keV energy band and the likely counterpart found searching in various on-line archives such 
as NED (NASA/IPAC Extragalactic Database), HEASARC (High Energy Astrophysics Science Archive Research Center) 
and SIMBAD (Set of Identifications, Measurements, and Bibliography for Astronomical Data). Following Harrison 
et al. (2003), we estimated the probability of detecting a 2--10 keV extragalactic source with flux greater 
$5 \times10^{-14}$ erg cm$^{-2}$ s$^{-1}$ by chance within a typical error circle of 3--5 arcmin radius to 
be negligible, i.e. close to zero over the entire sample.

Next, we analysed the spectra of the most likely X-ray counterparts, which have XRT data with
a signal to noise ratio above 5$\sigma$, in order to perform a reliable spectral analysis; for other
objects only flux information are provided.
Events for spectral analysis were extracted within a circular region of radius 
20$^{\prime \prime}$, centered on the source position; this region encloses about 90\% of the PSF at 1.5 
keV (see Moretti et al. 2004). The background was taken from source-free regions close to the X-ray source 
of interest, using circular regions with different radii in order to ensure an evenly sampled background. 
The source spectrum was then extracted from the corresponding event file using the {\sc XSELECT v.2.4} 
software and binned using {\sc grppha} in an appropriate way, so that the $\chi^{2}$ statistic could be 
applied. We used version v.011 of the response matrices and created the relative ancillary response file 
\textit{arf} using the task {\sc xrtmkarf v. 0.5.6}. The energy band used for the spectral analysis, 
performed with {\sc XSPEC} v. 11.3.2 (Arnaud 1996), depends on the statistical quality of the data and 
typically ranges from 0.3 to $\sim$8 keV.

For objects with more than one pointing, we summed together all the available observations and 
performed the spectral analysis of the average spectrum, unless flux variability was evident in the data;
in the first istance, we adopted, as our
basic model, a simple power law passing through Galactic absorption (Kalberla et al. 2005). If this baseline 
model was not sufficient to fit the data, we then introduced extra spectral components as required.
The results of this analysis are reported in Table~\ref{tab3}, where we list the Galactic
absorption, the column density in excess to this Galactic value, the power law photon index, the reduced 
$\chi^{2}$ of the best-fit model and the 2--10 keV flux.
All quoted errors correspond to 90$\%$ confidence level for a single
parameter of interest ($\Delta\chi^{2}=2.71$).
A more detailed description of the X-ray spectral analysis results is given in a dedicated 
section for each source.

\begin{table*}
\centering
\caption{Log of the \emph{Swift}/XRT observations used in this paper.}
\label{tab1}
\begin{tabular}{lccc}
\hline
\hline
IBIS source & ID & Obs. Date & Exposure \\
       &          &      & (s) \\
\hline
\hline
IGR J00465--4005           & 00038008001  & Nov 09, 2008 & 5015 \\
                           & 00038008002  & Nov 13, 2008 & 6656 \\
\hline
LEDA 96373                 & 00035621001  & Jun 04, 2006 & 8453 \\
                           & 00035621002  & Jun 19, 2006 & 3231 \\
                           & 00035621003  & Jun 22, 2006 & 2722 \\
\hline
IGR J07506--1547           & 00035349001  & Sep 22, 2006 & 1204 \\
                           & 00035349002  & Oct 12, 2006 & 8434 \\ 
\hline
1RXS J080114.6--462324     & 00262347001  & Feb 28, 2007 & 8084 \\
                           & 00262347002  & Mar 01, 2007 & 9028 \\
                           & 00262347003  & Mar 02, 2007 & 10846 \\
                           & 00262347004  & Mar 03, 2007 & 3607 \\
                           & 00262347005  & Mar 04, 2007 & 8164 \\
                           & 00262347006  & Mar 05, 2007 & 13593 \\
                           & 00262347008  & Mar 07, 2007 & 15228 \\
                           & 00262347009  & Mar 08, 2007 & 11510 \\
                           & 00262347010  & Mar 10, 2007 & 3561 \\
                           & 00262347011  & Mar 11, 2007 & 924 \\
                           & 00262347012  & Mar 12, 2007 & 5177 \\
                           & 00262347013  & Mar 13, 2007 & 4777 \\
                           & 00262347014  & Mar 14, 2007 & 2218 \\
\hline
IGR J08262+4051            & 00031311001  & Dec 16, 2008 & 5272 \\
                           & 00031311002  & Dec 17, 2008 & 4959 \\ 
                           & 00031311003  & Dec 23, 2008 & 8655 \\
                           & 00031311004  & Dec 30, 2008 & 5182 \\
                           & 00031311005  & Dec 31, 2008 & 980  \\
                           & 00031311006  & Jan 06, 2009 & 4039 \\
                           & 00031311007  & Jan 07, 2009 & 6201 \\
\hline
IGR J10447--6027           & 00031035001  & Dec 09, 2007 & 5055 \\ 
\hline
MCG+04--26--006            & 00037314001  & Jul 02, 2008 & 2362 \\
                           & 00037314002  & Jul 03, 2008 & 9341 \\
\hline
PKS 1143--696              & 00038056001  & Feb 05, 2009 & 3833 \\ 
                           & 00038056002  & Feb 28, 2009 & 5412 \\
\hline
IGR J12123--5802           & 00038059001  & Oct 17, 2008 & 6150 \\       
\hline
IGR J1248.2--5828        & 00038349001  & Dec 19, 2008 & 3771 \\
                           & 00038349002  & May 08, 2009 & 1867 \\
\hline
NGC 4748                   & 00035363001  & Dec 28, 2005 & 2277 \\
                           & 00035363002  & Jan 08, 2007 & 2605 \\
\hline 
IGR J13107--5626           & 00037092001  & Sep 23, 2007 & 13280 \\
\hline
IGR J14080--3023           & 00037384002  & Sep 17, 2008 & 7351 \\
                           & 00037384003  & Dec 16, 2008 & 1286 \\
                           & 00037384004  & Dec 17, 2008 & 9386 \\
\hline
IGR J17008--6425           & 00037053001  & Jun 10, 2007 & 5171  \\
\hline
IGR J17331--2406           & 00036121001  & Feb 27, 2007 & 6210 \\
                           & 00036121002  & Jul 01, 2007 & 368 \\
\hline
IGR J18134--1636           & 00037059001  & Jun 03, 2008 & 1224 \\
\hline
 IGR J18175--1530           & 00030991002  & Nov 03, 2007 & 1951 \\
\hline
\hline
\end{tabular}
\end{table*}
\begin{table*}
\centering
\setcounter{table}{0}
\caption{-- \emph{continued}}
\begin{tabular}{lccc}
\hline
\hline
IBIS source & ID & Obs. Date & Exposure \\
       &          &      & (s) \\
\hline
\hline
IGR J20569+4940            & 00038084001  & Feb 26, 2009 & 8453 \\
                           & 00038084002  & Mar 02, 2009 & 1465 \\
\hline 
1RXS J213944.3+595016      & 00035577001  & Mar 28, 2006 & 2388 \\
\hline
IGR J22234--4116           & 00037064001  & Jul 31, 2007 & 6710 \\
\hline
\hline
\end{tabular}
\end{table*}

\begin{table*}
\begin{center}
\caption{\emph{INTEGRAL}/IBIS position of the 20 selected sources 
and locations of the objects detected by XRT, within the 90\% and 99\% IBIS error circles, 
with relative counterparts. The error radii are given at 90\% confidence level.} 
\label{tab2}
\begin{tabular}{lccccc}
\hline
\hline
XRT source  & R.A.     &     Dec   &   error   &  Count rate(0.3--10 keV)  &  Counterpart$^{a}$  \\ 
  &   (J2000) &  (J2000) &   (arcsec)  & (10$^{-3}$ counts s$^{-1}$)  &        \\
\hline            
\multicolumn{6}{c}{IGR J00465--4005$^{b}$ (R.A.(J2000) = $00^{\rm h}46^{\rm m}27^{\rm s}.60$, Dec(J2000) =
$-40^\circ05^{\prime}13^{\prime \prime}.2$, error radius = 4$^{\prime}$.8)}\\
   &    &   &   &  &     \\
\#1 (in 90\%) & $00^{\rm h}46^{\rm m}20^{\rm s}.71$ & $-40^\circ05^{\prime}47^{\prime \prime}.3$ & 4.26
&  $15.2\pm1.7$ &  ESP 39607 \\
\hline
\multicolumn{6}{c}{LEDA 96373 (R.A.(J2000) = $07^{\rm h}26^{\rm m}27^{\rm s}.36$, Dec(J2000) =
$-35^\circ53^{\prime}31^{\prime \prime}.2$, error radius = 5$^{\prime}$.4)}\\
   &    &   &   &  &    \\
\#1 (in 99\%) & $07^{\rm h}26^{\rm m}10^{\rm s}.40$ & $-35^\circ47^{\prime}42^{\prime \prime}.3$ & 6.00 &
$2.13\pm0.64$ & 2MASS J07261422--3557401  \\
\#2 (in 90\%) & $07^{\rm h}26^{\rm m}14^{\rm s}.00$ & $-35^\circ57^{\prime}41^{\prime \prime}.9$ & 6.00 &
$2.52\pm0.69$ & 2MASS J07261006--3547403  \\
\#3 (in 90\%) & $07^{\rm h}26^{\rm m}26^{\rm s}.19$ & $-35^\circ54^{\prime}21^{\prime \prime}.3$ & 4.33 &
$10.7\pm1.3$ & LEDA 96373  \\
\#4 (in 99\%) & $07^{\rm h}26^{\rm m}51^{\rm s}.50$ & $-35^\circ48^{\prime}40^{\prime \prime}.3$ & 6.00 &
$8.26\pm1.20$ & 2MASS J07265165--3548412   \\
\hline
\multicolumn{6}{c}{IGR J07506--1547 (R.A.(J2000) = $07^{\rm h}50^{\rm m}42^{\rm s}.00$, Dec(J2000) =
$-15^\circ47^{\prime}34^{\prime \prime}.8$, error radius = 5$^{\prime}$.2)}\\
   &    &   &   &  &   \\
\#1 (in 99\%) & $07^{\rm h}50^{\rm m}19^{\rm s}.60$ & $-15^\circ51^{\prime}22^{\prime \prime}.2$ & 6.00
& $3.43\pm0.77$  & 2MASS J07501988--1551218    \\
\#2 (in 99\%) & $07^{\rm h}50^{\rm m}55^{\rm s}.10$ & $-15^\circ41^{\prime}44^{\prime \prime}.5$ & 6.00
&  $1.87\pm0.58$ &   USNO--B1.0	0743--0156384  \\
\hline
\multicolumn{6}{c}{1RXS J080114.6--462324 (R.A.(J2000) = $08^{\rm h}01^{\rm m}08^{\rm s}.16$, Dec(J2000) =
$-46^\circ22^{\prime}44^{\prime \prime}.4$, error radius = 3$^{\prime}$.6)}\\
   &    &   &   &  &    \\
\#1 (in 90\%) & $08^{\rm h}01^{\rm m}17^{\rm s}.09$ & $-46^\circ23^{\prime}26^{\prime \prime}.9$ & 3.75 &
$49.4\pm2.1$ &  1RXS J080114.6--462324   \\
\hline
\multicolumn{6}{c}{IGR J08262+4051$^{b}$ (R.A.(J2000) = $08^{\rm h}26^{\rm m}13^{\rm s}.44$, Dec(J2000) =
$+40^\circ51^{\prime}18^{\prime \prime}.0$, error radius = 4$^{\prime}$.8)}\\
   &    &   &   &  &    \\
\#1 (in 99\%) & $08^{\rm h}25^{\rm m}57^{\rm s}.70$ & $+40^\circ58^{\prime}21^{\prime \prime}.8$ & 
6.00 & $1.77\pm0.57$ & --  \\
\#2 (in 99\%) & $08^{\rm h}26^{\rm m}00^{\rm s}.00$ & $+40^\circ58^{\prime}55^{\prime \prime}.8$ & 
6.00 & $5.64\pm0.95$ &    MCG+07--18--001  \\
\#3 (in 90\%) & $08^{\rm h}26^{\rm m}17^{\rm s}.40$ & $+40^\circ47^{\prime}58^{\prime \prime}.1$ & 6.00 &
$2.95\pm0.75$ &  SDSS J082617.87+404758.6  \\
\#4 (in 99\%) & $08^{\rm h}26^{\rm m}46^{\rm s}.60$ & $+40^\circ55^{\prime}44^{\prime \prime}.0$ & 6.00 & 
$1.77\pm0.58$ & --   \\
\hline
\multicolumn{6}{c}{IGR J10447--6027 (R.A.(J2000) = $10^{\rm h}44^{\rm m}37^{\rm s}.20$, Dec(J2000) =
$-60^\circ25^{\prime}22^{\prime \prime}.8$, error radius = 4$^{\prime}$.1)}\\  
   &    &   &   &    &  \\
\#1 (in 90\%) & $10^{\rm h}44^{\rm m}51^{\rm s}.62$ & $-60^\circ25^{\prime}10^{\prime \prime}.6$ & 5.11 &
$6.31\pm1.30$ &  2MASS J10445192--6025115$^{c}$  \\
                  &    &    &   &        & 2MASS J10445200--6025102  \\
                  &    &    &   &        & 2MASS J10445118--6025121   \\ 
\hline
\multicolumn{6}{c}{MCG+04--26--006$^{b}$ (R.A.(J2000) = $10^{\rm h}46^{\rm m}53^{\rm s}.28$, Dec(J2000) =
$+25^\circ54^{\prime}10^{\prime \prime}.8$, error radius = 5$^{\prime}$.0)}\\
   &    &   &   &  &    \\
\#1 (in 90\%) & $10^{\rm h}46^{\rm m}42^{\rm s}.71$ & $+25^\circ55^{\prime}53^{\prime \prime}.2$ & 3.87 &
$27.3\pm2.0$ & MCG+04--26--006  \\
\hline
\multicolumn{6}{c}{PKS 1143--696 (R.A.(J2000) = $11^{\rm h}45^{\rm m}47^{\rm s}.76$, Dec(J2000) =
$-69^\circ53^{\prime}38^{\prime \prime}.4$, error radius = 4$^{\prime}$.1)}\\
   &    &   &   &  &     \\
\#1 (in 90\%) & $11^{\rm h}45^{\rm m}53^{\rm s}.73$ & $-69^\circ53^{\prime}59^{\prime \prime}.6$ & 3.60 &
$198.0\pm7.0$ & PKS 1143--696   \\
\hline
\multicolumn{6}{c}{IGR J12123--5802 (R.A.(J2000) = $12^{\rm h}12^{\rm m}15^{\rm s}.36$, Dec(J2000) =
$-58^\circ02^{\prime}56^{\prime \prime}.4$, error radius = 4$^{\prime}$.4)}\\
   &    &   &   &  &      \\
\#1 (in 90\%) & $12^{\rm h}12^{\rm m}25^{\rm s}.97$ & $-58^\circ00^{\prime}23^{\prime \prime}.1$ & 3.69 &
$85.4\pm4.0$ & 2MASS J12122623--5800204  \\
\#2 (in 90\%) & $12^{\rm h}12^{\rm m}32^{\rm s}.40$ & $-58^\circ06^{\prime}09^{\prime \prime}.5$ & 6.00 &
$2.56\pm0.79$ & --   \\
\hline
\multicolumn{6}{c}{IGR J1248.2--5828 (R.A.(J2000) = $12^{\rm h}47^{\rm m}46^{\rm s}.32$, Dec(J2000) =
$-58^\circ29^{\prime}13^{\prime \prime}.2$, error radius = 3$^{\prime}$.4)}\\
   &    &   &   &  &     \\
\#1 (in 99\%) & $12^{\rm h}47^{\rm m}38^{\rm s}.10$ & $-58^\circ24^{\prime}54^{\prime \prime}.3$ & 6.00 &
$4.65\pm1.40$ & HIP 62427   \\
\#2 (in 90\%) & $12^{\rm h}47^{\rm m}41^{\rm s}.57$ & $-58^\circ25^{\prime}53^{\prime \prime}.6$ & 3.94 &
$53.5\pm4.4$ & CCDM J12477--5826AB  \\
\#3 (in 90\%) & $12^{\rm h}47^{\rm m}57^{\rm s}.82$ & $-58^\circ29^{\prime}59^{\prime \prime}.1$ & 4.02 &
$45.6\pm4.1$ & 2MASX J12475784--5829599  \\
\hline
\multicolumn{6}{c}{NGC 4748$^{b}$ (R.A.(J2000) = $12^{\rm h}52^{\rm m}12^{\rm s}.00$, Dec(J2000) =
 $-13^\circ25^{\prime}48^{\prime \prime}.0$, error radius = 5$^{\prime}$.5)}\\
   &    &   &   &  &      \\
\#1 (in 90\%) & $12^{\rm h}52^{\rm m}12^{\rm s}.28$ & $-13^\circ24^{\prime}54^{\prime \prime}.0$ & 3.63 &
$316.0\pm13.0$ & NGC 4748  \\
\hline
\multicolumn{6}{c}{IGR J13107--5626 (R.A.(J2000) = $13^{\rm h}10^{\rm m}40^{\rm s}.56$, Dec(J2000) =
$-56^\circ26^{\prime}52^{\prime \prime}.8$, error radius = 4$^{\prime}$.0)}\\
   &    &   &   &  &    \\
\#1 (in 90\%) & $13^{\rm h}10^{\rm m}37^{\rm s}.27$ & $-56^\circ26^{\prime}56^{\prime \prime}.7$ & 4.43 &
$4.51\pm0.74$ & 2MASX J13103701--5626551   \\
\hline
\multicolumn{6}{c}{IGR J14080--3023$^{b}$ (R.A.(J2000) = $14^{\rm h}08^{\rm m}02^{\rm s}.16$, Dec(J2000) =
$-30^\circ23^{\prime}31^{\prime \prime}.2$, error radius = 3$^{\prime}$.8)}\\
   &    &   &   &  &   \\
\#1 (in 90\%) & $14^{\rm h}08^{\rm m}06^{\rm s}.57$ & $-30^\circ23^{\prime}52^{\prime \prime}.6$ & 3.55 &
$252.0\pm6.1$ &  2MASX J14080674--3023537  \\ 
\hline
\hline
\end{tabular}
\end{center}
\end{table*}
\begin{table*}
\setcounter{table}{1}
\caption{-- \emph{continued}}
\begin{center}
\begin{tabular}{lcccccc}
\hline
\hline
XRT source  & R.A.     &     Dec   &   error   &  Count rate(0.3--10 keV)  &  Counterpart$^{a}$  \\ 
  &   (J2000) &  (J2000) &   (arcsec)  & (10$^{-3}$ counts s$^{-1}$)  &        \\

\hline
\multicolumn{6}{c}{IGR J17008--6425$^{b}$ (R.A.(J2000) = $17^{\rm h}00^{\rm m}25^{\rm s}.44$, Dec(J2000) =
$-64^\circ24^{\prime}10^{\prime \prime}.8$, error radius = 5$^{\prime}$.0)}\\
   &    &   &   &  &   \\
\#1 (in 99\%) & $16^{\rm h}59^{\rm m}34^{\rm s}.10$ & $-64^\circ21^{\prime}29^{\prime \prime}.7$ & 6.00 & 
$21.1\pm2.6$ & --   \\
\#2 (in 99\%) & $16^{\rm h}59^{\rm m}50^{\rm s}.10$ & $-64^\circ28^{\prime}35^{\prime \prime}.9$ & 6.00 & 
$4.81\pm1.20$ & --    \\
\#3 (in 90\%) & $16^{\rm h}59^{\rm m}56^{\rm s}.70$ & $-64^\circ24^{\prime}16^{\prime \prime}.5$ & 6.00 &
$5.48\pm1.10$ & 2MASS J16595624--6424198   \\ 
\#4 (in 90\%) & $17^{\rm h}00^{\rm m}18^{\rm s}.30$ & $-64^\circ21^{\prime}15^{\prime \prime}.5$ & 6.00 &  
$6.94\pm1.50$ & --   \\
\#5 (in 99\%) & $17^{\rm h}00^{\rm m}19^{\rm s}.90$ & $-64^\circ29^{\prime}34^{\prime \prime}.3$ & 6.00 &
$44.7\pm3.3$ & --    \\
\#6 (in 90\%)  & $17^{\rm h}00^{\rm m}30^{\rm s}.80$ & $-64^\circ23^{\prime}41^{\prime \prime}.2$ & 6.00 & 
$12.8\pm1.9$ &  USNO--B1.0 0256--0660529   \\
\#7 (in 90\%) & $17^{\rm h}00^{\rm m}45^{\rm s}.10$ & $-64^\circ25^{\prime}42^{\prime \prime}.2$ & 6.00 &
$6.13\pm1.30$ & USNO--B1.0 0255--0654745 	 \\ 
\#8 (in 99\%) & $17^{\rm h}01^{\rm m}06^{\rm s}.20$ & $-64^\circ18^{\prime}59^{\prime \prime}.7$ & 6.00 &
$11.1\pm1.2$ & 2MASS J17010602--6419055   \\
\#9 (in 99\%) & $17^{\rm h}01^{\rm m}15^{\rm s}.20$ & $-64^\circ23^{\prime}43^{\prime \prime}.2$ & 6.00 &
$6.32\pm1.40$ & --   \\ 
\#10 (in 99\%) & $17^{\rm h}01^{\rm m}17^{\rm s}.90$ & $-64^\circ28^{\prime}52^{\prime \prime}.5$ & 6.00 & 
$6.42\pm1.40$ & 2MASS J17011786-6428481    \\
\#11 (in 99\%) & $17^{\rm h}01^{\rm m}25^{\rm s}.10$ & $-64^\circ28^{\prime}48^{\prime \prime}.2$ & 6.00 & 
$3.17\pm1.00$ & --    \\
\#12 (in 99\%) & $17^{\rm h}01^{\rm m}25^{\rm s}.90$ & $-64^\circ26^{\prime}38^{\prime \prime}.6$ & 6.00 &
$34.9\pm2.9$ & --     \\
\#13 (in 99\%) & $17^{\rm h}01^{\rm m}32^{\rm s}.70$ & $-64^\circ22^{\prime}10^{\prime \prime}.7$ & 6.00 &
$5.15\pm1.30$ & --    \\
\#14 (in 99\%) & $17^{\rm h}01^{\rm m}33^{\rm s}.90$ & $-64^\circ25^{\prime}19^{\prime \prime}.7$ & 6.00 & 
$8.98\pm1.60$ & --  \\
\#15 (in 99\%) & $17^{\rm h}01^{\rm m}36^{\rm s}.60$ & $-64^\circ23^{\prime}08^{\prime \prime}.8$ & 6.00 & 
$4.81\pm1.30$ & 2MASS J17013585--6423117   \\
\#16 (in 99\%) & $17^{\rm h}01^{\rm m}37^{\rm s}.30$ & $-64^\circ23^{\prime}53^{\prime \prime}.3$ & 6.00 & 
$26.6\pm2.8$ & --  \\
\hline
\multicolumn{6}{c}{IGR J17331--2406 (R.A.(J2000) = $17^{\rm h}33^{\rm m}13^{\rm s}.44$, Dec(J2000) =
$-24^\circ08^{\prime}34^{\prime \prime}.8$, error radius = 1$^{\prime}$.4)}\\
   &    &   &   &  &   \\
 -- & -- & -- & -- & -- & --  \\
\hline
\multicolumn{6}{c}{IGR J18134--1636 (R.A.(J2000) = $18^{\rm h}13^{\rm m}26^{\rm s}.88$, Dec(J2000) =
$-16^\circ37^{\prime}15^{\prime \prime}.6$, error radius = 3$^{\prime}$.7)}\\
   &    &   &   &  &     \\
 -- & -- & -- & -- & -- & --  \\
\hline
\multicolumn{6}{c}{IGR J18175--1530 (R.A.(J2000) = $18^{\rm h}17^{\rm m}42^{\rm s}.24$, Dec(J2000) =
$-15^\circ28^{\prime}19^{\prime \prime}.2$, error radius = 4$^{\prime}$.7)}\\
   &    &   &   &  &    \\
 -- & -- & -- & -- & -- & --  \\
\hline
\multicolumn{6}{c}{IGR J20569+4940 (R.A.(J2000) = $20^{\rm h}56^{\rm m}41^{\rm s}.04$, Dec(J2000) =
$+49^\circ41^{\prime}02^{\prime \prime}.4$, error radius = 4$^{\prime}$.2)}\\
   &    &   &   &  &    \\
\#1 (in 90\%)& $20^{\rm h}56^{\rm m}42^{\rm s}.64$ & $+49^\circ40^{\prime}08^{\prime \prime}.9$ & 3.55 &
$265.0\pm6.6$ & 1RXS J205644.3+494011  \\
\hline
\multicolumn{6}{c}{1RXS J213944.3+595016 (R.A.(J2000) = $21^{\rm h}39^{\rm m}42^{\rm s}.72$, Dec(J2000) =
$+59^\circ49^{\prime}37^{\prime \prime}.2$, error radius = 4$^{\prime}$.5)}\\
   &   &   &   &   &    \\
\#1 (in 90\%) & $21^{\rm h}39^{\rm m}44^{\rm s}.72$ & $+59^\circ50^{\prime}15^{\prime \prime}.7$ & 3.80 & 
$139.0\pm8.5$ & 2MASS J21192912+3332566  \\
\hline
\multicolumn{6}{c}{IGR J22234--4116$^{b}$ (R.A.(J2000) = $22^{\rm h}23^{\rm m}24^{\rm s}.00$, Dec(J2000) =
$-41^\circ15^{\prime}36^{\prime \prime}.0$, error radius = 4$^{\prime}$.1)}\\
   &   &   &   &   &    \\
\#1 (in 99\%) & $22^{\rm h}23^{\rm m}01^{\rm s}.30$ & $-41^\circ18^{\prime}28^{\prime \prime}.3$ & 6.00 & 
$4.87\pm1.1$ & --  \\
\#2 (in 90\%) & $22^{\rm h}23^{\rm m}10^{\rm s}.90$ & $-41^\circ18^{\prime}59^{\prime \prime}.2$ & 6.00 & 
$4.74\pm1.0$ & 1RXS J222313.2--411923  \\
\#3 (in 90\%) & $22^{\rm h}23^{\rm m}11^{\rm s}.00$ & $-41^\circ15^{\prime}23^{\prime \prime}.0$ & 6.00 &
$5.48\pm1.1$ & -- \\ 
\#4 (in 90\%) & $22^{\rm h}23^{\rm m}26^{\rm s}.30$ & $-41^\circ19^{\prime}43^{\prime \prime}.8$ & 6.00 &  
$3.78\pm0.95$ & --    \\
\#5 (in 90\%) & $22^{\rm h}23^{\rm m}33^{\rm s}.70$ & $-41^\circ18^{\prime}13^{\prime \prime}.9$ & 6.00 &
$4.19\pm1.0$ & --  \\
\#6 (in 99\%) & $22^{\rm h}23^{\rm m}45^{\rm s}.00$ & $-41^\circ12^{\prime}10^{\prime \prime}.6$ & 6.00 &
$8.97\pm1.4$ & SUMSS J222344--411150  \\ 
\#7 (in 90\%)  & $22^{\rm h}23^{\rm m}46^{\rm s}.30$ & $-41^\circ15^{\prime}41^{\prime \prime}.1$ & 6.00 & 
$3.81\pm0.98$ & --   \\
\#8 (in 90\%) & $22^{\rm h}23^{\rm m}46^{\rm s}.50$ & $-41^\circ15^{\prime}10^{\prime \prime}.8$ & 6.00 &
$7.96\pm1.4$ & --  \\ 
\#9 (in 99\%) & $22^{\rm h}23^{\rm m}47^{\rm s}.90$ & $-41^\circ17^{\prime}27^{\prime \prime}.5$ & 6.00 & 
$6.11\pm1.2$ & 2MASS J22234827--4117276 \\
\hline
\hline
\end{tabular}
\begin{list}{}{}
$^{a}$ From NED, HEASARC and SIMBAD;\\
$^{b}$ This source is located at high Galactic latitude ($|b| > 10^{\circ}$);\\
$^{c}$ The better localisation obtained by \emph{Chandra} indicates this source as the likely counterpart 
of the IBIS detection (see text). 
\end{list}
\end{center}
\end{table*}

\section{Notes on individual sources}
In the following, results on each individual source are presented. All sources discussed in this section 
appear in the fourth IBIS catalogue (Bird et al. 2009) either as new detections or as already
reported sources; if no mention to previous references is given, then the source is a new \emph{INTEGRAL} 
detection.
 
\subsection*{\bf IGR J00465--4005} 
\begin{figure}
\includegraphics[width=1.0\linewidth]{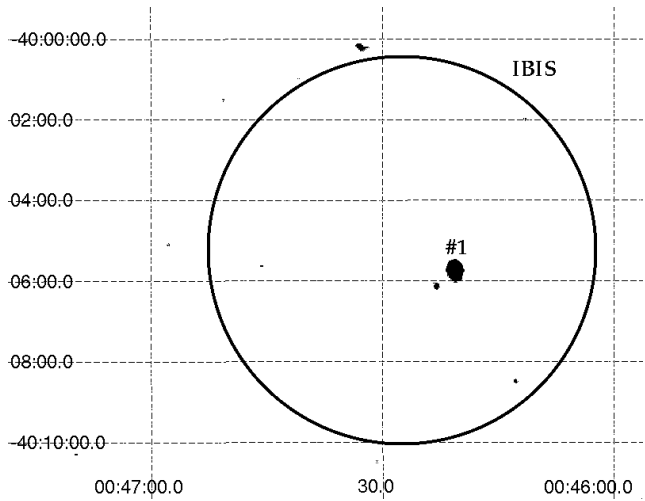}
\caption{XRT 0.3--10 keV image of the region surrounding IGR J00465--4005.
Source \#1 is the only source detected by XRT within the IBIS uncertainty (black circle).}
\label{fig1}
\end{figure}

This IBIS source is located at high Galactic latitude ($b \sim -77^{\circ}$), which already suggests an 
extragalactic nature.
According to an archival search, a ROSAT Faint X-ray source, 1RXS J004638.2--400921, is present within 
the IBIS positional uncertainty, but its location is known with no great accuracy (15 arcsec radius). 
There is only one XRT source detected within the IBIS error circle at $\sim$9$\sigma$ confidence level 
in the longest ($\sim$6.7 ks) XRT observation (see Figure~\ref{fig1}) and it coincides with the ROSAT 
object, which is now located  with an accuracy of few arcsec. This source  has a radio counterpart in the NVSS 
(NRAO VLA Sky Survey, Condon et al. 1998) with a 20 cm flux of $41.5\pm1.3$ mJy and 
in the SUMMS (The Sydney University Molonglo Sky Survey, Mauch et al. 2003)  with a 36 cm flux of 
$60.3\pm2.3$ mJy. Furthermore, the XRT detection coincides with the galaxy ESP 39607 reported in NED 
as having a redshift of $z=0.226$. The source is listed in the United States Naval Observatory 
(USNO--B1.0, Monet et al. 2003) catalogue with $R$, $B$ magnitudes of 17.3 and 17.8 respectively, 
while in the 2MASS (2 Micron All Sky Survey, Skrutskie et al. 2006) survey it is reported as having
near-infrared magnitudes $J=16.379\pm0.129$, $H=15.844\pm0.164$ and $K=14.845\pm0.133$.

The fit with our basic model does not satifactorily reproduce the data ($\chi^2/{\nu} = 60.8/19$) 
because of the presence of excess emission below 2 keV. To account for this extra feature, we adopted a 
double power law model, with the primary component absorbed by intrinsic absorption and the secondary 
component, having the same photon index of the primary one, passing only through the Galactic column 
density: with this model we found a fit improvement corresponding to a $\Delta\chi^2/{\nu} = 49.5/2$.
As a result (see Table~\ref{tab3}), we obtain  an intrinsic absorption of $N_{\rm H}$ 
$\sim$$2.4\times10^{23}$ cm$^{-2}$, a steep photon index ($\Gamma$ $\sim$2.5) and a 2--10 keV flux of
$\sim$$1.2\times 10^{-12}$ erg cm$^{-2}$ s$^{-1}$. The X-ray spectroscopy also indicates a flux 
variability by a factor 1.3 between the two XRT observations.

The overall characteristics of IGR J00465--4005 strongly 
indicates that ESP 39607 is an active galaxy at intermediate redshift, while the absorption measured
in the X-ray spectrum further indicates that it is an absorbed AGN. This hypothesis is now confirmed by 
dedicated optical spectroscopy (Masetti et al. 2009).

\subsection*{\bf LEDA 96373}     
\begin{figure}
\includegraphics[width=1.0\linewidth]{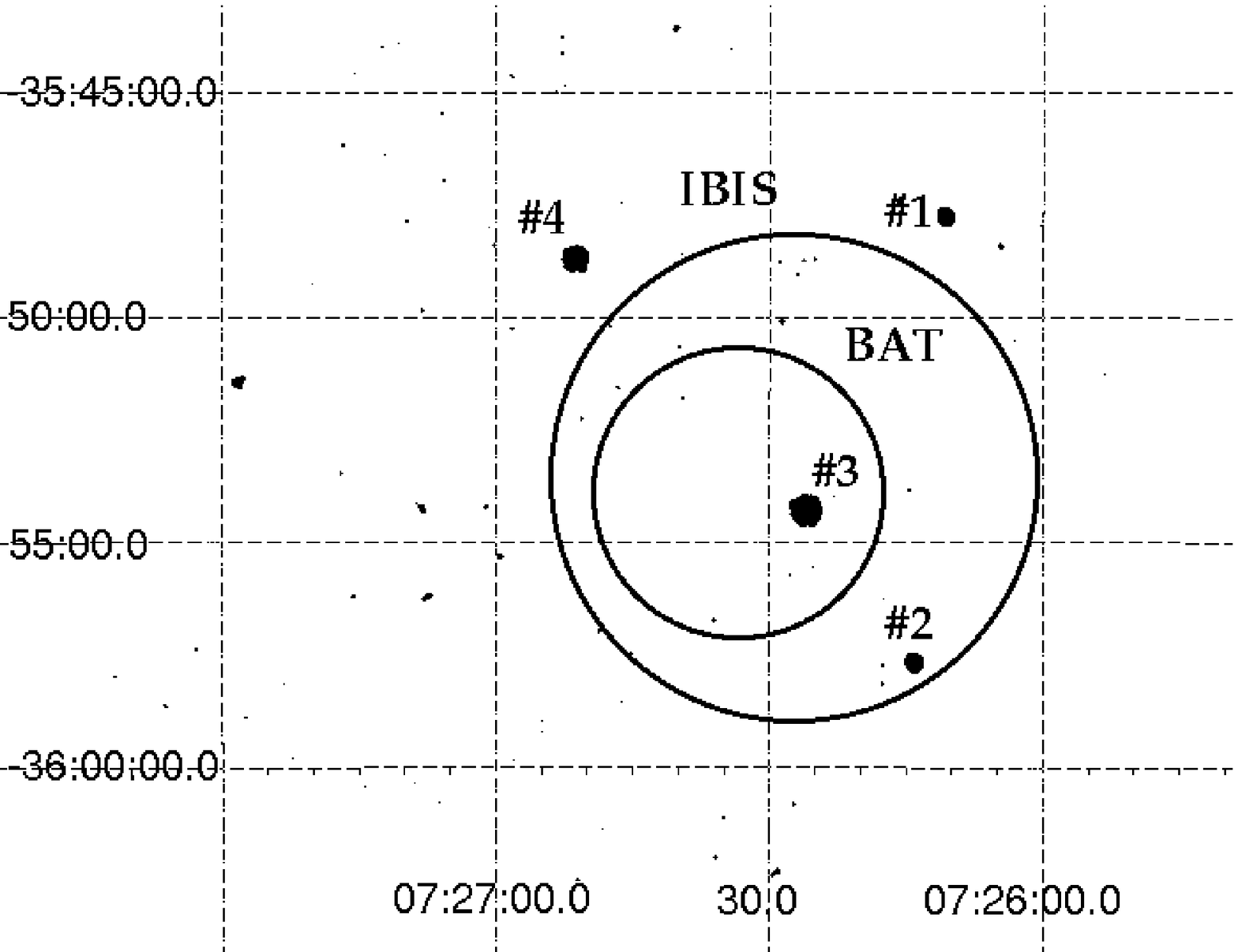}
\caption{XRT 0.3--10 keV image of the region surrounding LEDA 96373.
The larger and smaller black circles represent the IBIS and BAT positions and uncertainties, respectively.
Also plotted are the positions of the four sources detected by XRT within the 90\% (\#2 and \#3) and 99\% 
(\#1 and \#4) IBIS error circles.}
\label{fig2}
\end{figure}

This source, known also as IGR J07264--3553, was first observed at high energies during the all-sky hard 
X-ray IBIS survey (Krivonos et al. 2007) and recently reported in the Palermo \emph{Swift}/BAT survey 
(Cusumano et al. (2009); both works associated it to LEDA 96373 (also 2MASX J07262635--3554214) a galaxy
classified in NED as a Seyfert 2 at $z=0.0294$. However, X-ray follow-up observations with XRT indicate 
the presence of more possible counterparts in the IBIS error circle as listed in Table~\ref{tab2} and 
shown in Figure~\ref{fig2}): four 
X-ray sources are visible of which two (\#1 and \#4) are located within the 99\% error radius and the other 
two (\#2 and \#3) are instead found within the 90\% positional uncertainty. Sources \#1 and \#2 are 
weak and soft being detected mostly below 3 keV, but the other two objects are brighter and of comparable 
intensity; object \#4 is mostly emitting soft X-rays. Source \#3 is indeed the galaxy LEDA 96373 proposed 
as the likely counterpart of the hard X-ray emitter. The combined use of both IBIS and BAT error circles 
(see Figure~\ref{fig2}) clearly indicates that only this source is detected by both instruments, thus 
confirming previous associations.
The source is yet another AGN found close to the Galactic plane; it is an IRAS source (IRAS 07245--3548),
as well as a relatively bright radio object (NVSS J070726--355422) having a 20 cm flux of $171.7\pm5.2$ mJy.

Despite the low statistical quality of the data, the basic model is not a good fit, as
excess emission is observed below 2 keV. Also in this case, a double power law model is a better fit 
($\Delta\chi^2/{\nu} = 7.5/2$); we find an intrinsic column density of
$\sim$$7\times10^{22}$ cm$^{-2}$ and a photon index $\Gamma$ $\sim$2.5 (see Table~\ref{tab3})
compatible within uncertainties with the canonical AGN value. The source seems to vary by a factor of 2 
within a time-scale of few days, with an average 2--10 keV flux of $\sim$$4.2\times10^{-13}$ erg cm$^{-2}$ 
s$^{-1}$. The X-ray spectral characteristics, reported in this work for the first time, are compatible with 
the AGN type 2 nature of this source.

\subsection*{\bf IGR J07506--1547} 
\begin{figure}
\includegraphics[width=1.0\linewidth]{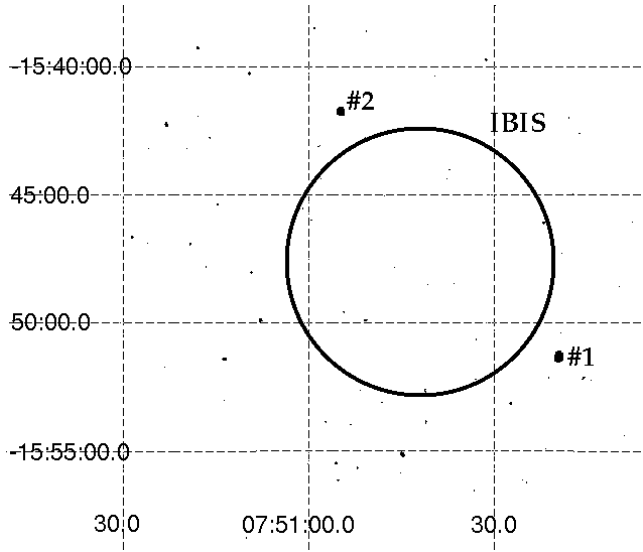}
\caption{XRT 0.3--10 keV image of the region surrounding IGR J07506--1547.
Sources \#1 and \#2, the only objects detected by XRT, are located within the 99\% IBIS error circle.}
\label{fig3}
\end{figure}

This source was first reported in the second IBIS catalogue (Bird et al. 2006),
but not confirmed in the third survey (Bird et al 2007) most likely
because it went into a quiescent period which prevented detection at later stages
of the analysis. Indeed, it was recovered in the fourth catalogue thanks to the bursticity analysis,
which provided a clear source detection when period of quiescence were removed (see Bird et al. 2009).    
There are two X-ray observations available for this field, but unfortunately
no source was detected within the 90\% IBIS uncertainty during either XRT pointing.
In the longer observation ($\sim$8.3 ks) two objects (\#1 and \#2 in Table~\ref{tab2}) are 
revealed, but they lie within the 99\% IBIS error circle (see Figure~\ref{fig3}); furthermore, 
neither are detected when extracting the XRT image above 3 keV. Source \#1 has a counterpart in the 
USNO--B1.0 catalogue located at RA(J2000) = $07^{\rm h}50^{\rm m}55^{\rm s}.26$, 
Dec(J2000) = $-15^\circ41^{\prime}43^{\prime \prime}.9$ (with magnitudes $R$ $\sim$20.4);
this object has no infrared counterpart in the 2MASS catalogue.
For source \#2, we find a counterpart in a USNO--B1.0 object located at RA(J2000) = $07^{\rm h}50^{\rm 
m}19^{\rm s}.95$, Dec(J2000) = $-15^\circ51^{\prime}21^{\prime \prime}.9$ (with magnitudes $R$ $\sim$15),
also listed in the 2MASS survey, with magnitudes $J=13.732\pm0.029$,
$H=13.435\pm0.027$ and $K=13.319\pm0.051$. Lacking further information, we cannot establish in more detail 
their nature. However, their faintness in X-rays\footnote{By assuming a power law model, we find a 2--10 
keV flux of $\sim$7$\times10^{-14}$ erg 
cm$^{-2}$ s$^{-1}$ for source \#1 and $\sim$3$\times10^{-13}$ erg cm$^{-2}$ s$^{-1}$ for source \#2.},
soft spectrum and location with respect to the IBIS error circle suggest that their association with IGR 
J07506--1547 is unlikely. On the other hand, the lack of a bright counterpart is consistent with the strong 
variable X-ray emission of this IBIS object. It is evident that it will be very difficult to catch the soft 
X-ray counterpart of IGR J07506--1547 using ``standard'' follow-up observations unless a well defined 
monitoring strategy can be conceived.
 
\subsection*{\bf 1RXS J080114.6--462324}

\begin{figure}
\includegraphics[width=1.0\linewidth]{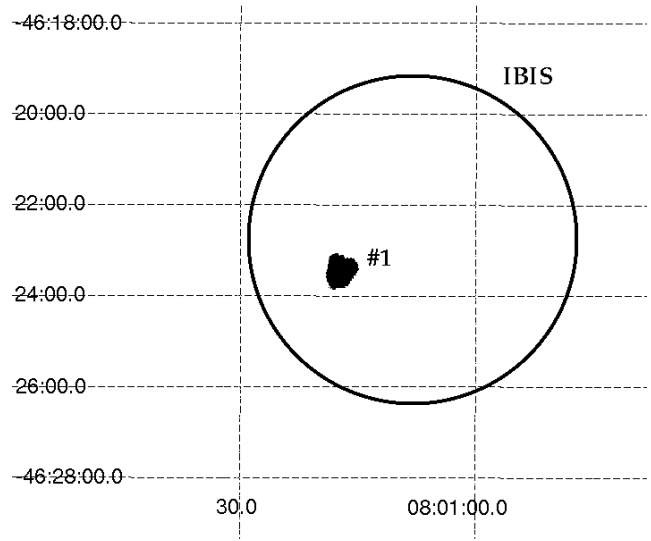}
\caption{XRT 0.3--10 keV image of the region surrounding 1RXS J080114.6--462324.
Within the IBIS error circle, XRT detects only one object (source \#1).}
\label{fig4}
\end{figure}

Within the fourth IBIS catalogue this source is labelled as variable with a small bursticity value (Bird 
et al. 2009). The ROSAT Faint Survey source is clearly the counterpart as it lies very close to the IBIS 
best-fit position; however, its positional accuracy is too poor (23 arcsec) to allow optical follow-up 
observations. Fortunately, this sky region was observed repeatedly by XRT as follow-up observations of 
GRB\,070227 so that monitoring of the IBIS/ROSAT detection was also possible. The analysis of the XRT 
images provides a better position for the ROSAT detection and confirms its associations with the IBIS 
object (see Figure~\ref{fig4}), being this X-ray source also detected above 3 keV. This object is 
also reported in the 
\emph{XMM-Newton} Slew Survey (XMMSL1 J080117.3--462328) with a positional accuracy comparable to that of 
XRT and a 0.2--12 keV flux of $4.1 \times 10^{-12}$ erg cm$^{-2}$ s$^{-1}$, compatible with the range of 
values measured in the same energy band during the XRT observations ($\sim$($1.6-5.1) \times 10^{-12}$ erg 
cm$^{-2}$ s$^{-1}$). Within the XRT error circle we find a USNO--B1.0 object located at RA(J2000) = 
$08^{\rm h}01^{\rm m}16^{\rm s}.88$, Dec(J2000) = $-46^\circ23^{\prime}27^{\prime \prime}.9$, with
magnitudes $R$ $\sim$16.3, which also belongs to the 2MASS survey with
magnitudes $J = 14.442\pm0.043$, $H = 14.032\pm0.051$ and $K = 13.744\pm0.059$; 
this source is just compatible with the \emph{XMM-Newton} Slew positional uncertainty and furthermore 
is reported as a star with a proper motion (46.3 (R.A.) and 18.5 (Dec) mas/yr) in the
PPM--Extended catalogue (R\"{o}ser et al. 2008). 

A simple power law does not reproduce the X-ray data ($\chi^2/{\nu} = 216.4/108$) again because of the 
presence of soft emission below 2 keV; modelling this component with a blackbody emission provides an  
improvement of the fit ($\Delta\chi^2/{\nu} = 109.9/2$). 
The best-fit results, reported in Table~\ref{tab3}, are a photon index
$\Gamma$ $\sim$1.6 and a blackbody temperature of $kT$ $\sim$100 eV. No extra absorption in excess to the 
Galactic value is required by the data. 
As mentioned above, comparing different XRT pointings, we find significant changes in flux (by a factor 
of $\sim$3.2), but not in the spectral shape (although a small variation is observed in the power law photon 
index). This observational evidence confirms that the source variability also observed in the IBIS 
data (Bird et al. 2009) is a typical signature of this source. The X-ray 
spectral behaviour, as well as the short-term variability (of the order of few days) and the observation 
of proper motion, argue in favour of a Galactic source, while the source location slightly above the 
Galactic plane ($b$ $\sim$$8^{\circ}$) suggests that it might be a cataclysmic variable. Indeed, recent 
optical follow-up observations confirm this suggestion (Masetti et al. 2009).

\subsection*{\bf IGR J08262+4051} 

\begin{figure}
\includegraphics[width=1.0\linewidth]{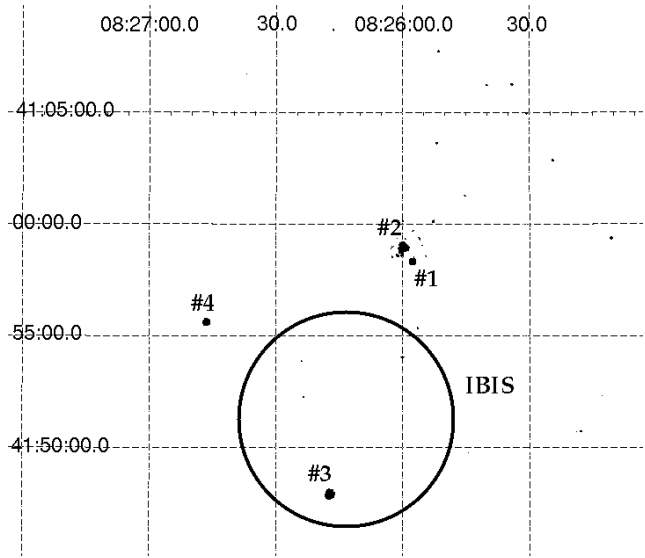}
\caption{XRT 0.3--10 keV image of the region surrounding IGR J08262+4051.
There is only one source (\#3) detected within the 99\% IBIS error circle; the 
other three objects (\#1, \#2 and \#4) are located within the 99\% IBIS uncertainty.}
\label{fig5}
\end{figure}

This source was found only analysing \emph{INTEGRAL} data of Revolution 314, 
suggesting a variable nature. In this case, XRT detects four sources, of which three are located within 
the 99\% and one within the 90\% IBIS error circle (see Table~\ref{tab2} and Figure~\ref{fig5}), 
but only in the longer 
($\sim$8.6 ks) XRT pointing (see Table~\ref{tab1}). The brightest object (\#2), detected at $\sim$6$\sigma$ 
confidence level in the 0.3--10 keV energy band but not above 3 keV, coincides with the bright galaxy 
MCG+07--18--001, also listed in the 2MASS Extended catalogue (2MASX J08260056+4058514) and previously 
detected as a ROSAT faint object (1RXS J082609.2+405808); this source with a redshift of $z=0.0574$ is a 
member of the galaxy group SDSS--C4--DR3 3247 (from NED). Within the XRT positional uncertainty, we find  
a NVSS radio source (NVSS J082600+405850) having a 20 cm flux of $14.0\pm1.0$ mJy, most probably associated 
to MCG+07--18--001. The X-ray spectrum of this galaxy is well described (see Table~\ref{tab3}) by a 
power law having a photon index $\Gamma$$\sim$2 and a 2--10 keV flux of 
$\sim$1.4$\times10^{-13}$ erg cm$^{-2}$ s$^{-1}$.

The second object in brightness is source \#3, revealed by XRT at $\sim$4$\sigma$ confidence 
level: it has an optical counterpart in an USNO--B1.0 object located at RA(J2000) =
$08^{\rm h}26^{\rm m}17^{\rm s}.88$, Dec(J2000) = $+40^\circ47^{\prime}59^{\prime \prime}.0$
(with magnitude $R=19.6 - 20.3$) and it is also listed in the Sloan Digital Sky Survey (SDSS) as a quasar 
candidate (SDSS J082617.87+404758.6, Gordon et al. 2004) with redshift $z=0.975$. No infrared counterpart 
has been found in the 2MASS catalogue. The statistical quality of the XRT data of this source is too poor 
for a proper characterisation of the X-ray spectrum, allowing only a flux estimate in the 2--10 keV
energy band of $\sim$5$\times10^{-14}$ erg cm$^{-2}$ s$^{-1}$ to be made.

For the two remaining objects (\#1 and \#4), within their XRT positional uncertainty, 
we do not find any catalogued object in the various databases quaried; their 2--10 keV flux is very weak
being around $\sim$2$\times10^{-14}$ erg cm$^{-2}$ s$^{-1}$.

The variable nature of IGR J08262+4051 suggests that any of these four X-ray objects could be the
counterpart if the source was caught in a quiescent state during the XRT observation. 
Given the extragalactic nature of two of these XRT detections, the presence of a galaxy group in the 
region and the high Galactic latitude of the IBIS source ($b$ $\sim$$34^{\circ}$) it is likely that also 
the other two sources are of similar nature, i.e. galaxies. We therefore conclude that IGR J08262+4051 
is most likely an AGN of still unclassified type and that dedicated optical follow-up observations of 
the four X-ray detections can determine their class and define which is associated to the IBIS source.

\subsection*{\bf IGR J10447--6027} 
\begin{figure}
\includegraphics[width=1.0\linewidth]{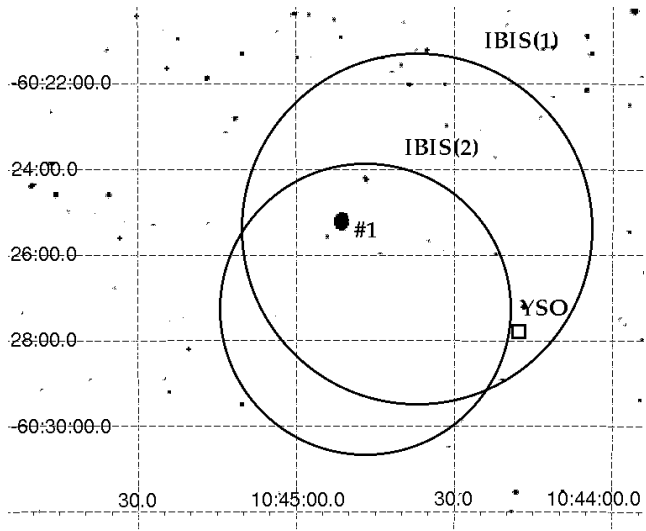}
\caption{XRT 0.3--10 keV image of the region surrounding IGR J10447--6027.
The larger (IBIS(1)) and smaller (IBIS(2)) black circles represent the IBIS positions and uncertainties as 
reported by Bird et al. (2009) and Leyder, Walter \& Rauw (2008), respectively. The only X-ray source 
detected by XRT is labelled as \#1. Also plotted (black box) is the position of the young stellar
object (YSO, IRAS 10423--6011) proposed by Leyder et al. (2008) as the counterpart of IGR J10447--6027.}
\label{fig6}
\end{figure}

This source was discovered by Leyder, Walter \& Rauw (2008) during the analysis of the region surrounding 
Eta Carinae and associated to a young stellar object (YSO, IRAS 10423--6011). The characteristics of the 
IBIS spectrum could be interpreted either as evidence for a new High Mass X-ray Binary, or as a signature 
of accretion and/or particle acceleration in the YSO. The analysis of the XRT image, shown in 
Figure~\ref{fig6}, clearly indicates the presence of an X-ray source within the IBIS error circles 
reported by Bird et al. (2009) (larger black circle, IBIS(1)) and Leyder et al. (2008) (smaller 
black circle, IBIS(2)), while the lack 
of an XRT detection at around the YSO position leads us to discard the association between the YSO and 
the IBIS object proposed by Leyder et al. (2008).
The only X-ray source visible in the XRT image is detected at $\sim$4.9$\sigma$ confidence level in the 
0.3--10 keV energy band (see Table~\ref{tab2}). A further more detailed analysis shows that this source
is not detected below 5 keV, but its significance in the hard energy range (5--10 keV) is 
4.4$\sigma$ ($4.41\pm1.00$ counts s$^{-1}$) indicating a very hard spectrum and confirming the association 
to the IBIS source. Within the XRT uncertainty we find three objects belonging to the 2MASS survey (see 
Table~\ref{tab2}); these 
infrared/XRT sources have no optical counterparts in the USNO--B1.0 catalogue or at other wavelengths.
Fortunately, a \emph{Chandra} observation of this region provides further constraints on the location and 
positional uncertainty of the XRT source (RA(J2000) = $10^{\rm h}44^{\rm m}51^{\rm s}.99$,
Dec(J2000) = $-60^\circ25^{\prime}12^{\prime \prime}.3$, less than 1
arcsec error radius (Fiocchi private comunication) and helps pinpoint only one 2MASS/infrared 
counterpart of the X-ray/gamma-ray object, i.e. the one reported in Table~\ref{tab2}. This source has 
magnitudes $J$ and $H$ of $\sim$15 and magnitude $K$ of $\sim$14, which combined with the optical non 
detection in USNO--B1.0 implies a quite red object.

The X-ray spectroscopy (see Table~\ref{tab3}), although of low statistical quality, indicates an absorbed
power law ($N_{\rm H}$ $\sim$$2\times10^{23}$ cm$^{-2}$, $\Gamma = 1.8$ frozen), in line with the source 
reddening. Clearly, only infrared follow-up observations, particularly spectroscopic measurements, could 
possibly shed light on the nature of this intriguing source.

\subsection*{\bf MCG+04--26--006} 
\begin{figure}
\includegraphics[width=1.0\linewidth]{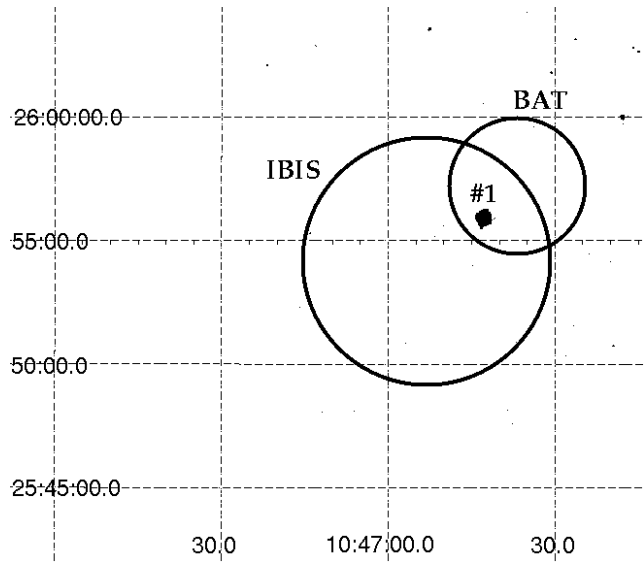}
\caption{XRT 0.3--10 keV image of the region surrounding MCG+04--26--006.
There is only one source (\#1) detected by XRT within both the IBIS and BAT error circles.}
\label{fig7}
\end{figure}

This source also appears in the recent \emph{Swift}/BAT survey of Cusumano et al. (2009) and it is
associated with MCG+04--26--006, also named UCG 05881, reported in NED as a galaxy at $z=0.020$. 
The region is however crowded with objects and MCG+04--26--006 itself belongs to a compact group of 
galaxies, so that it is not clear if we are seeing one galaxy or a cluster of galaxies.
Within the combined IBIS and BAT positional uncertainties, there is only one X-ray source detected at 
$\sim$14$\sigma$ confidence level (see Figure~\ref{fig7}).
It coincides with MCG+04--26--006 which is reported both in the USNO--B1.0 catalogue
with magnitude $R=9.85-10.04$, in the 2MASS Extended object list (2MASX J10464247+2555540) and as an IRAS 
source. The XRT localization is also compatible with a NVSS radio source (NVSS J104642+255552) having a 20 
cm flux of $10.7\pm0.5$ mJy.

The X-ray spectrum indicates (see Table~\ref{tab3}) an absorbed ($N_{\rm H}$ $\sim$$1.2 \times10^{23}$
cm$^{-2}$) steep ($\Gamma$ $\sim$3) power law having a 2--10 keV flux of $\sim$$2\times10^{-12}$ erg
cm$^{-2}$ s$^{-1}$. It is possible that thermal emission from the galaxy group contaminates
the X-ray spectrum providing a softer spectrum than generally observed in AGN.
However, the fit is equally acceptable ($\chi^2/{\nu} = 12.2/13$) if we fix the photon index to the 
AGN canonical value of 1.8;
in this case the absorption is slightly less ($\sim$$7.5\times10^{22}$ cm$^{-2}$), but still in 
excess to the Galactic value. The source shows a flux variability of a factor of 2, during XRT pointings, 
with a time-scale of a day, although no variation in spectral shape is observed; this suggests that AGN 
emission is likely to be dominating any thermal emission if present. Just on the basis of these results, 
we can conclude that the IBIS source is probably extragalactic, most likely a type 2 AGN as 
described. Fortunately, 
the source is listed in the SDSS catalogue and so an optical spectrum is available to confirm our findings: 
the source is classified as a LINER (Low Ionization Nuclear Emission Region, Masetti et al. 2009).
The absorption observed suggests that it could be a type 2 LINER, i.e. one without the broad H$\alpha$ 
component. This would then indicate that the source is powered by an absorbed active nucleus rather than 
by starburst activity.

\begin{table*}
\begin{center}
\caption{\emph{Swift}/XRT spectral analysis results of the averaged spectra. Frozen parameters are 
written in square brackets; errors are given at the 90\% confidence level.}
\label{tab3}
\begin{tabular}{lccccc}
\hline
\hline
 Source & $N_{\rm H}(Gal)$ & $N_{\rm H}$ & $\Gamma$ & $\chi^2/\nu$  & $F_{\rm (2-10~keV)}$ \\
  & ($10^{22}$ cm$^{-2}$) &  ($10^{22}$ cm$^{-2}$)   &  &  &   ($10^{-11}$ erg cm$^{-2}$ s$^{-1}$) \\
\hline
\hline
IGR J00465--4005$^{a}$ (\#1)  &  0.0344 & $24.1^{+9.1}_{-6.1}$ & $2.49^{+0.43}_{-0.26}$ & 11.3/17 
& $0.12\pm0.01$ \\ 
\hline
LEDA 96373$^{a}$ (\#3) & 0.260   & $7.0^{+9.9}_{-3.8}$ & $2.55^{+0.70}_{-0.37}$ & 3.5/7 
& $0.042\pm0.004$ \\
\hline
1RXS J080114.6--462324$^{b}$ (\#1)  & 0.221 & --  &  $1.55^{+0.11}_{-0.14}$ & 106.5/106
& $0.16\pm0.01$ \\
\hline
IGR J08262+4051 (\#2)         & 0.0405& --  &  $2.07^{+0.45}_{-0.43}$ & 7.3/14
& $0.014\pm0.002$ \\   
\hline
IGR J10447--6027 (\#1)      & 1.27   & $25.6^{+29.0}_{-13.0}$ & [1.8]  & 3.4/4 
& $0.105\pm0.021$ \\
\hline
MCG+04--26--006 (\#1)      & 0.0251   & $12.4^{+4.6}_{-3.8}$ & $3.16^{+1.23}_{-1.07}$ & 8.2/12
& $0.22\pm0.01$ \\
\hline
PKS 1143--696  (\#1)       & 0.162    &       --         &  $1.74\pm0.10$ & 71.8/53 
& $0.54\pm0.02$ \\
\hline
IGR J12123--5802 (\#1)     & 0.325    &            --         &  $1.26^{+0.18}_{-0.16}$ & 19.7/18
& $0.45\pm0.02$ \\
\hline
IGR J1248.2--5828 (\#3)    & 0.297    & $0.92^{+0.67}_{-0.82}$ &  $0.86^{+0.76}_{-0.70}$ & 9.4/14
& $0.41\pm0.03$ \\
\hline
NGC 4748 (\#1)       & 0.0352   &        --         &  $2.20\pm0.11$ & 33.3/40
& $0.34\pm0.01$ \\
\hline
IGR J13107--5626 (\#1)     & 0.244    & $39.3^{+24.4}_{-13.3}$ & 
[1.8]         & 7.5/9
& $0.11\pm0.02$ \\
\hline
IGR J14080--3023$^{c}$ (\#1) & 0.0362 &        --          & $1.41\pm0.06$ & 144.0/134
& $0.64\pm0.01$  \\
\hline
IGR J20569+4940 (\#1)        & 1.00     & $0.53^{+0.18}_{-0.16}$ & $2.32^{+0.18}_{-0.17}$ & 86.3/87
& $1.19\pm0.02$ \\
\hline
1RXS J213944.3+595016 (\#1)    & 0.568        & $0.35^{+0.37}_{-0.21}$ & $1.99^{+0.40}_{-0.35}$ & 
13.4/15   & $0.59\pm0.06$ \\
\hline
\hline
\end{tabular}
\begin{list}{}{}
$^{a}$ Best-fit model requires a second power law component, having the same photon 
index of the primary absorbed power law, and passing only through the Galactic column density;\\
$^{b}$ Best-fit model includes a black-body component with a $kT = 110^{+9}_{-7}$ eV
to account for the excess observed below 2 keV;\\
$^{c}$ Best-fit model includes a black-body component with a $kT = 81^{+6}_{-5}$ eV
to account for the excess observed below 2 keV.
\end{list}
\end{center}
\end{table*}

\subsection*{\bf PKS 1143--696}
\begin{figure}
\includegraphics[width=1.0\linewidth]{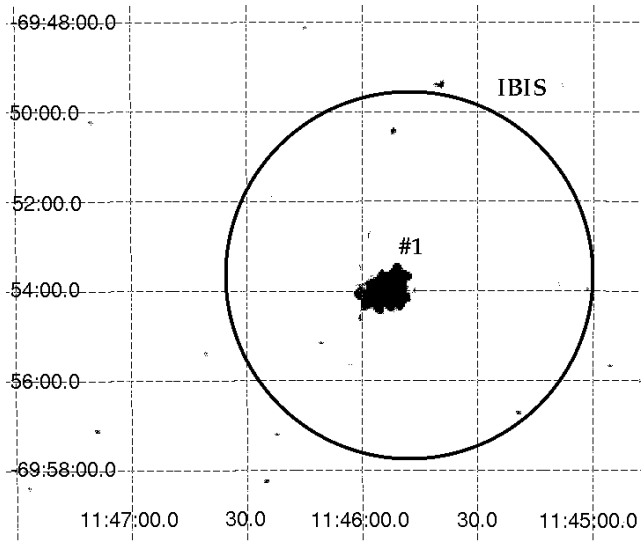}
\caption{XRT 0.3--10 keV image of the region surrounding PKS 1143--696.
Source \#1 is the only X-ray source detected by XRT within the IBIS uncertainty.}
\label{fig8}
\end{figure}

This source, first reported by Krivonos et al. (2007) in their all-sky hard X-ray survey 
(Krivonos et al. 2007), is classified in SIMBAD/NED as a possible quasar; the object is also listed in the 
ROSAT Bright Survey (1RXS J114553.1--695349) and in many radio catalogues, giving support to the association 
with the IBIS source. PKS 1143--696 was reported in the fourth catalogue (Bird et al. 2009) as it was detected
by mean of the bursticity analysis, which provides a clear detection. This implies that the source is 
variable above 10 keV. The XRT image provides only one source detected at $\sim$28$\sigma$ confidence level 
(see Figure~\ref{fig8}) and coincident with PKS 1143--696. Within the XRT positional uncertainty, 
there is only one object in the USNO--B1.0 catalogue having magnitude $R$ $\sim$15.7 and in the 2MASS 
survey with magnitudes $J = 14.828\pm0.040$, $H = 13.928\pm0.050$ and $K = 12.952\pm0.038$.

The X-ray spectrum is well fitted with an unabsorbed power law having a photon index $\Gamma$ $\sim$1.7 and 
a 2--10 keV flux of $\sim$$5.4 \times10^{-12}$ erg cm$^{-2}$ s$^{-1}$.
During the XRT observations, the source showed variability in flux (by a factor of 1.2 within a time-scale 
of a month), but not in spectral shape.

To characterise the source in the radio band, we have used SpecFind (Vollmer et al 2005),
which is a tool able to cross-identify radio sources in various catalogues
on the basis of a self-consistent spectral index as well position. This allows us
to combine data at different frequencies and to estimate the source radio spectrum
($S_{\nu} \propto \nu^\alpha$); for this object the spectrum has a slope of --0.2
indicative of a compact radio source. Clearly, PKS 1143--696 is an AGN with many X-ray and radio properties 
resembling those of a QSO; follow-up optical spectroscopy should be able to classify this source more 
properly.

\subsection*{\bf IGR J12123--5802} 

\begin{figure}
\includegraphics[width=1.0\linewidth]{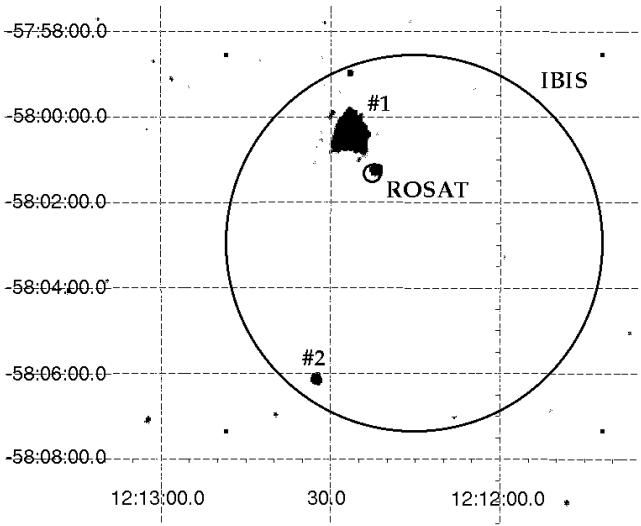}
\caption{XRT 0.3--10 keV image of the region surrounding IGR J12123--5802.
The larger black circle represents the IBIS position and uncertainty, while the two X-ray sources detected 
within it are labelled as \#1 and \#2. Also plotted is the position (smaller black circle) of a ROSAT 
Bright Survey source (1RXS J121222.7--580118) located $\sim$1$^{\prime}$ away from source \#1.}
\label{fig9}
\end{figure}

Another interesting case is that of IGR J12123--5802. Within the IBIS uncertainty,  
an archival search indicates the presence of a ROSAT Bright Survey source (1RXS J121222.7-580118)
located with an accuracy of 12 arcsec. XRT reveals instead two X-ray 
sources (\# 1 and \#2 as shown in Figure~\ref{fig9}) detected at $\sim$21$\sigma$ and 
$\sim$3.2$\sigma$, respectively. Source \#1 has a counterpart in the USNO--B1.0 catalogue located at 
RA(J2000) = $12^{\rm h}12^{\rm m}26^{\rm s}.22$, Dec(J2000) = $-58^\circ00^{\prime}20^{\prime \prime}.5$
(with magnitude $R=14.88-16.45$), also listed in the 2MASS survey,
with magnitudes $J = 15.435\pm0.061$, $H = 15.178\pm0.106$ and $K = 15.101\pm0.177$.

No counterpart in any database is found for source \#2, which is detected up to 4 keV.

At $\sim$1$^{\prime}$ away from source \#1 we find 
the ROSAT source 1RXS J121222.7--580118 (see Figure~\ref{fig9}): it is detected just below $\sim$3$\sigma$ 
confidence level and only up to 3 keV; it has has a  2--10 keV flux of $\sim$$3 
\times10^{-13}$ erg cm$^{-2}$ s$^{-1}$ assuming a power law model with 
the photon index frozen to 1.8. A variable behaviour of this source could be a viable way to explain 
why it is listed in the ROSAT Bright catalogue, but it is faint during the XRT observation.
Since IGR J12123--5802 is listed as persistent in the fourth IBIS catalogue, it cannot be associated with
1RXS J121222.7--580118 nor to source \#2 given its weak  X-ray flux and soft spectrum; thus, source \#1 
becomes the natural counterpart of the IBIS detection.

The X-ray spectrum of this source is well described by a simple power law having a flat photon index ($\Gamma$
$\sim$1.3) and a 2--10 keV flux of $\sim$$4.5\times10^{-12}$ erg cm$^{-2}$ s$^{-1}$ (see Table~\ref{tab3}). 
Despite being able to pinpoint the X-ray and optical counterpart of IGR J12123--5802 in source \#1, 
we cannot infer anything about its classification. Also in this case, optical spectroscopy will be 
necessary to identify this high energy source.

\subsection*{\bf IGR J1248.2--5828} 

\begin{figure}
\includegraphics[width=1.0\linewidth]{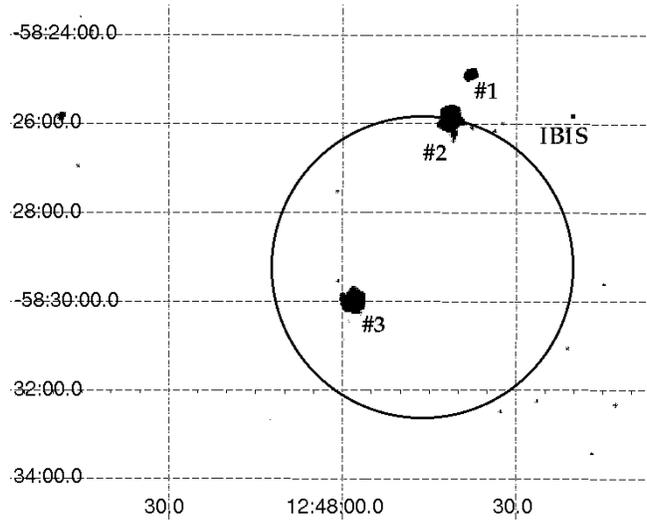}
\caption{XRT 0.3--10 keV image of the region surrounding IGR J1248.2--5828.
The white circle represents the 90\% IBIS error circle, while the two X-ray sources detected by XRT within 
and at the border of it are labelled as \#2 and \#3. Source \#1 is the fainter 
source detected by XRT within the 99\% IBIS error circle.}
\label{fig10}
\end{figure}

In this case, we find three X-ray sources located within the IBIS uncertainty (see Table~\ref{tab2} 
and Figure~\ref{fig10}):
one (source \#1) is detected within the 99\% IBIS error circle, while the second (source
\#2) and the third (source \#3) ones are located at the border and inside of the 90\% IBIS uncertainty, 
respectively. From a detailed analysis of the XRT image of the longest observation
($\sim$3.7 ks), source \#1 turns out to be the faintest object of the three detections, being detected at 
$\sim$3.3$\sigma$ over the entire XRT energy band. Its position is compatible 
with an object classified as a star (HIP 62427) in SIMBAD.
Source \#2 is brighter in soft X-rays having a count rate of ($53.5\pm4.4)\times10^{-3}$ counts s$^{-1}$ 
even compared to source \#3 (($45.6\pm4.1)\times10^{-3}$ counts s$^{-1}$).
Its XRT position is coincident with CCDM J12477--5826AB, classified as a double or
multiple star in SIMBAD, and it is also compatible with a ROSAT Faint Survey source (1RXS 
J124742.1--582544), 
which is still unclassified. The third object (\#3), detected at 11$\sigma$ confidence level in the
0.3--10 keV energy band and still revealed above 3 keV, has a counterpart in a
USNO--B1.0 source
located at RA(J2000) = $12^{\rm h}47^{\rm m}57^{\rm s}.85$, Dec(J2000) = $-58^\circ29^{\prime}59^{\prime 
\prime}.9$ (with magnitude $R=14.61-15.13$), and it is also listed in the 2MASS Extended survey (2MASX 
J12475784--5829599). The object location is also compatible with a radio source belonging to
the MGPS--2 (Molonglo Galactic Plane Survey 2nd Epoch Compact Source, Murphy et al. 2007) 
catalogue, with a 36 cm flux of $16.5\pm1.1$ mJy.

The XRT spectra of source \#1 and \#2 are very soft as no emission is detected above 3 keV; both of them 
are fitted by a thermal bremsstrahlung model with $kT$ $\sim$0.4 keV and a 2--10 keV flux of $\sim$$2.1
\times10^{-15}$ erg cm$^{-2}$ s$^{-1}$ and $\sim$$2.5 \times10^{-14}$ erg cm$^{-2}$ s$^{-1}$, 
respectively.
Their positional coincidence with stellar objects suggests that the X-ray emission comes from a stellar
corona and is not related to the IBIS source. 

The X-ray spectrum of source \#3 (see Table~\ref{tab3}) is instead well described by a power law model 
having a flat photon index 
($\Gamma$ $\sim$0.9) and a 2--10 keV flux of $\sim$$4\times 10^{-12}$ erg cm$^{-2}$ s$^{-1}$. By assuming 
a photon index of 1.8, the fit is still acceptable ($\Delta\chi^2/{\nu} = 13.0/15$) and provides a high 
intrinsic absorption of $\sim$$2\times 10^{22}$ cm$^{-2}$. During the XRT observations this source shows a 
flux variability of a factor of 2.5 on a time-scale of a few months.

The X-ray brightness and hard spectrum of source \#3 argue in favour of its association with IGR 
J1248.2--5828; furthemore, the information collected on this X-ray counterpart, i.e. its association to a 
2MASS Extended object and to a radio source, lead us to propose IGR J1248.2--5828 as an AGN behind the 
Galactic plane. This suggestion has recently been confirmed by optical follow-up observations (Masetti et 
al. 2009).

\subsection*{\bf NGC 4748}

\begin{figure}
\includegraphics[width=1.0\linewidth]{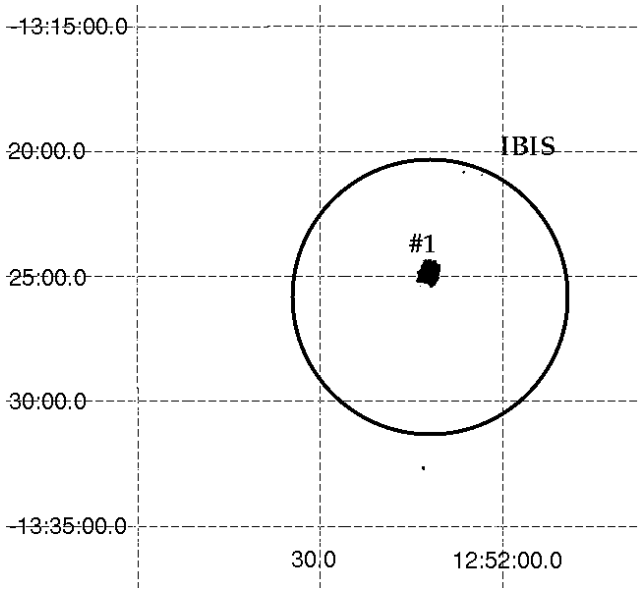}
\caption{XRT 0.3--10 keV image of the region surrounding NGC 4748.
XRT detects only one X-ray object (source \#1) within the IBIS error circle.}
\label{fig11}
\end{figure}

Within the IBIS error circle (see Figure~\ref{fig11}), we find only one X-ray source at 
$\sim$24$\sigma$ confidence level, which
coincides with NGC 4748, classified as a Seyfert 1 galaxy in NED with redshift
$z=0.01463$. It is reported in the 2MASS Extended catalogue (2MASX J12521245--1324528) and also in the 
ROSAT Bright Survey (1RXS J125212.5--132450). The XRT localization is also compatible with a
NVSS radio source (NVSS J125212--132452) having a 20 cm flux of $14.0\pm0.6$ mJy.
A more carefully analysis of the source references in the literature shows that NGC 4748 is better 
classified as a Narrow Line Seyfert galaxy (V\'eron-Cetty, V\'eron \& Gon\c calves 2001).

The X-ray spectroscopy (see Table~\ref{tab3}) indicates an unabsorbed power law having a photon index 
$\Gamma$ $\sim$2, a 2--10 keV flux of $\sim$$3.3 \times 10^{-12}$ erg cm$^{-2}$ s$^{-1}$ and a flux 
variation of a factor of $\sim$1.5 on time-scale of years during the XRT observations, i.e. fully 
compatible with its Seyfert class.

\subsection*{\bf IGR J13107--5626} 
\begin{figure}
\includegraphics[width=1.0\linewidth]{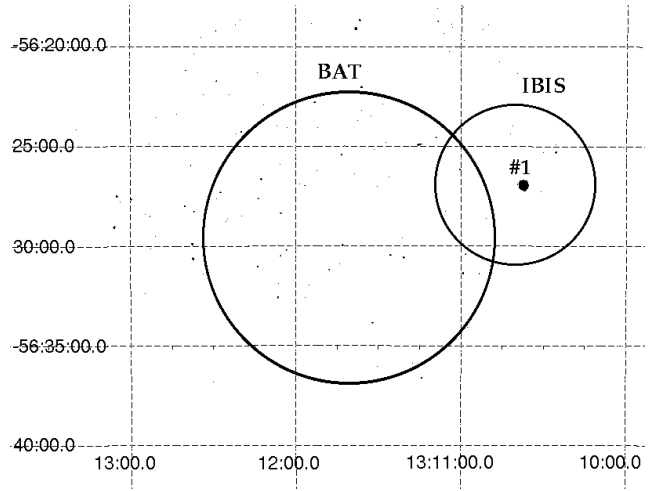}
\caption{XRT 0.3--10 keV image of the region surrounding IGR J13107--5626.
The larger and smaller black circles represent the IBIS and BAT position and uncertainty, respectively, 
while the black box shows the location of the 2MASS Extended object (2MASX J13103701--5626551) proposed as 
counterpart.}
\label{fig12}
\end{figure}

The only source detected in the entire XRT field of view, at $\sim$6$\sigma$ 
confidence level, lies right in the middle of the IBIS error circle. It has a counterpart in a 
2MASS Extended object (2MASX J13103701--5626551), which is
classified as a galaxy of unknown type in NED. It is reported in the USNO--B1.0 
catalogue with magnitudes $R=16.57-17.24$ and $B=16.57-17.24$. 
The XRT position is also compatible with a radio source belonging to
the MGPS--2 catalogue with a 36 cm flux of $35.4\pm1.6$ Jy. 
As can be seen in Figure~\ref{fig12}, although 
the IBIS uncertainty partially overlaps the BAT error circle of Swift J1312.1--5631 (Tueller et al. 2009),
the two sources seem to be uncorrelated. 2MASX J13103701--5626551, which is proposed by Tueller et al.
(2009) as the counterpart of the \emph{Swift} object, is well located within the IBIS uncertainty, but it 
is $\sim$$1^{\prime}.9$ away from the border of the BAT error circle and $\sim$$9^{\prime}.4$ from
the BAT centroid position. Clearly, while the associations between the 2MASS Extended galaxy and the 
IBIS high energy emitter is evident, the connection with the BAT detection is less convincing. 
We note that this source is not reported in the most recent 39 month \emph{Swift}/BAT catalogue by 
Cusumano et al. (2009).

The XRT source is faint (2--10 keV flux of $\sim$$1\times 10^{-12}$ erg cm$^{-2}$ s$^{-1}$) and
because of the low quality of the XRT data we fixed the power law photon index to 1.8, finding an
absorption in excess to the Galactic one of $\sim$$4\times 10^{23}$ cm$^{-2}$ 
(see Table~\ref{tab3}). 

On the basis of the multiwaveband properties (emission in radio and X-rays and extension in infrared) 
we conclude that IGR J13107--5626 is an AGN, possibly absorbed.

\subsection*{\bf IGR J14080--3023}
\begin{figure}
\includegraphics[width=1.0\linewidth]{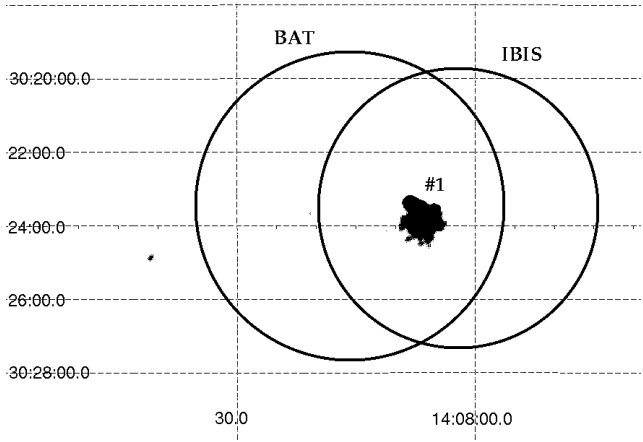}
\caption{XRT 0.3--10 keV image of the region surrounding IGR J14080--3023. Only one X-ray object
(source \#1) is detected by XRT within both the IBIS and BAT uncertainties.}
\label{fig13}
\end{figure}

This source also appears in the recent \emph{Swift}/BAT survey of Cusumano et al. (2009) where it is 
associated with a 2MASS Extended object (2MASX J14080674--3023537) classified as a Seyfert 1 at 
$z=0.024$ in SIMBAD. Indeed, the only object detected by XRT within the IBIS error circle (see 
Figure~\ref{fig13}), 
at $\sim$41$\sigma$ confidence level, coincides with the Seyfert galaxy.
This source is reported in the USNO--B1.0 catalogue with magnitude $R=13.72-14.17$ and it is also listed
in the \emph{XMM-Newton} Slew Survey (XMMSL1 J140806.7--302348) with a positional uncertainty greater 
($\sim$8.8 arcsec) than that of XRT (see Table~\ref{tab2}) and a 0.2--12 keV flux of $3.2 \times 10^{-12}$ 
erg cm$^{-2}$ s$^{-1}$, lower than the range measured during the XRT pointings ($\sim$($0.9-1.3) \times 
10^{-11}$ erg cm$^{-2}$ s$^{-1}$) in the same energy band. 
Here too a closer look at the literature (V\'eron-Cetty \& V\'eron 2001)
indicates that the source is better classified as a Seyfert of intermediate (1.5) type.

The fit of its X-ray spectrum, reported here for the first time, with the basic model does not reproduce 
satisfactorily the data ($\chi^2/{\nu} = 338.6/135$). The presence of soft emission below 2 keV is well 
described by a blackbody component, which provides a fit improvement corresponding to a 
$\Delta\chi^2/{\nu} = 196.4/1$.
This best-fit model provides a photon index $\Gamma$ $\sim$1.4 and a blackbody temperature $kT$ $\sim$81 
eV (see Table~\ref{tab3}). The data do not require extra absorption in excess to the Galactic one.
The source shows a slight flux variability (by a factor of 1.3), with a time-scale of a few months, but 
no variations in spectral shape are observed. The X-ray spectral characteristics are 
consistent with the AGN type 1 classification proposed for this source.

\subsection*{\bf IGR J17008--6425}
\begin{figure}
\includegraphics[width=1.0\linewidth]{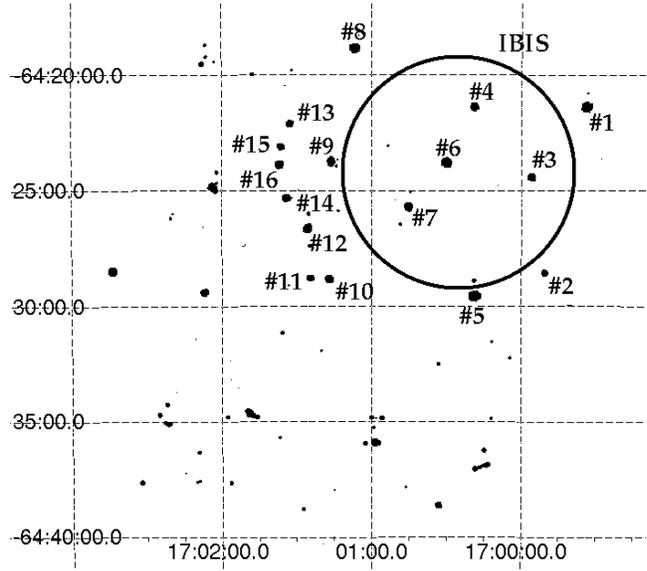}
\caption{XRT 0.3--10 keV image of the region surrounding IGR J17008--6425.
XRT detectes four X-ray source (\#3, \#4, \#6 and \#7) within the 90\% IBIS error circle, while
the remaining objects shown in the figure are located with the 99\% IBIS uncertainty.}
\label{fig14}
\end{figure}

This object, first reported in the third IBIS catalogue (Bird et al. 2007) 
as persistent, is detected in the fourth survey by mean of the bursticity 
analysis, thus indicating a variable behaviour on long time-scale. The analysis of the XRT 
image reveals the presence of 16 X-ray detections, of 
which 4 are located within the 90\% and 12 within the 99\% IBIS error circle (see Table~\ref{tab2} and 
Figure~\ref{fig14}); all these sources disappear above 3 keV.
If we combine the positional uncertainties
of the third and fourth survey, it is possible to limit the number of possible associations
within the 90\% error circles to sources \#6 and \#7. Within their 
XRT uncerainty, there is only one object in the USNO--B1.0 catalogue 
having magnitude $R$ $\sim$18.3 and $R$ $\sim$18.8, respectively. Of these two objects,
source \#6 is the most promising as it is slightly brighter than the companion.

Apart from source \#1, all the remaining 13 X-ray detections are located within the common area covered by 
the 99\% IBIS error circles of the third and fourth surveys.
Searching in various databases, we find a possible counterpart only for four objects (see 
Table~\ref{tab2}). 
Source \#3 has a counterpart in a USNO B--1.0 catalogue located at RA(J2000) =
$16^{\rm h}59^{\rm m}56^{\rm s}.22$, Dec(J2000) = $-64^\circ24^{\prime}19^{\prime \prime}.7$
(with magnitude $R$ $\sim$17.0--17.8), also belonging to the 2MASS survey with magnitudes 
$J = 16.548\pm0.121$, $H = 16.159\pm0.169$ and $K = 15.385$. We also find a radio source, located 
$\sim$2.3 arcmin from the centroid of the XRT position, belonging to the SUMSS catalogue
(SUMSS J170136--643051) having a 36 cm flux of $7.6\pm0.9$ mJy.
The XRT position of source \#8 is compatible with those of a USNO B--1.0 object located at RA(J2000) =
$17^{\rm h}01^{\rm m}06^{\rm s}.08$, Dec(J2000) = $-64^\circ19^{\prime}05^{\prime \prime}.5$
(with magnitude $R$ $\sim$16.2--16.7), also belonging to the 2MASS survey with magnitudes
$J = 15.831\pm0.060$, $H = 15.507\pm0.110$ and $K = 15.307\pm0.166$. 
Within the XRT positional uncertainty of source \#10, there is a USNO B--1.0 catalogue located at 
RA(J2000) = $17^{\rm h}01^{\rm m}17^{\rm s}.89$, Dec(J2000) = $-64^\circ28^{\prime}48^{\prime 
\prime}.0$ (with magnitude $R$ $\sim$16.0--16.3), which also belongs to the 2MASS survey with 
magnitudes $J = 14.966\pm0.043$, $H = 14.444\pm0.044$ and $K = 14.261\pm0.061$. 
There are two USNO--B1.0 objects within the XRT uncertainty of source \#15, the brightest of 
the two being reported as a 2MASS object having magnitudes $J = 13.254\pm0.030$, $H = 12.966\pm0.034$ and
$K = 12.921\pm0.038$. Unfortunately, the information collected about these objects cannot help us in 
pinpointing the true association.

Furthermore, the faintness of all these X-ray detections prevents us from performing a proper 
spectral analysis and allows only an
estimate of the 2--10 keV flux, which lies in the range $\sim$$(0.4-3)\times10^{-13}$ erg cm$^{-2}$ 
s$^{-1}$. 

In conclusion, based on the above information, no firm identification can be provided for IGR J17008--6425, 
although its location at high Galactic latitudes ($b \sim -13^{\circ}$) suggests an extragalactic 
nature.

\subsection*{\bf IGR J17331--2406}  
\begin{figure}
\includegraphics[width=1.0\linewidth]{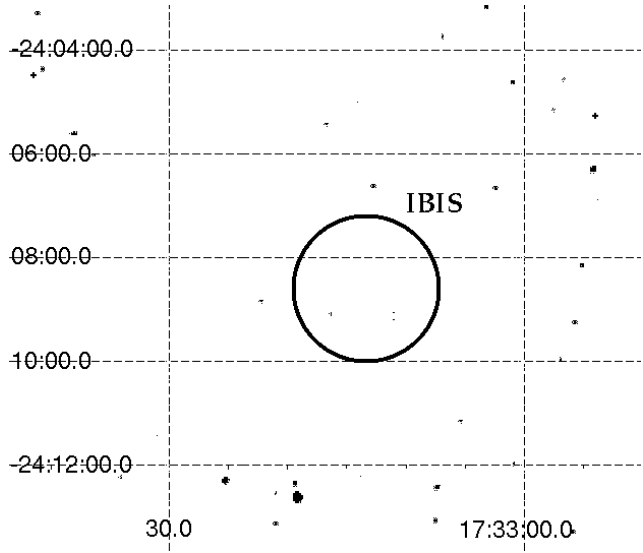}
\caption{XRT 0.3--10 keV image of the region surrounding IGR J17331--2406.
No X-ray source is found by XRT within the IBIS uncertainty.}
\label{fig15}
\end{figure}

Very little is known about this source other than it is a transient caught in outburst in
2004 (Lutovinov et al. 2004). No optical counterpart has so far been suggested and a \emph{Chandra} 
observation of the region did not detect any X-ray source (Tomsick et al 2008). 
In the fourth IBIS catalogue, the source is found through the bursticity analysis, which is an indication of 
variability. No XRT detection is found within the IBIS error circle (see Figure~\ref{fig15}), 
highlighting the difficulties of accurately locating sources that are either transient 
or very variable. Our upper limit on the source X-ray flux is $\sim$$6\times 10^{-14}$ erg cm$^{-2}$ 
s$^{-1}$ (0.3--10 keV) higher than the \emph{Chandra} one which is in the range ($1-1.5)\times 10^{-14}$ 
erg cm$^{-2}$ s$^{-1}$) (Tomsick et al. 2008). This is clearly a source which spend 
considerable time in quiescence (it is possibly a very faint 
transient) and will therefore be very difficult to classify.

\subsection*{\bf IGR J18134--1636}
This is a persistent source, first reported by Bird et al (2006) in the second IBIS catalogue.
Up to very recently no counterpart was reported or suggested for this object, but
a \emph{Chandra} observation of the region revealed the presence of an X-ray source  
named CXOU J181328.0--163548: this source has no optical or IR counterpart and an absorbed X-ray spectrum 
with a 0.3--10 keV flux of $\sim$$3\times 10^{-12}$ erg cm$^{-2}$ s$^{-1}$ (Tomsick et al. 2009). 
 
\begin{figure}
\includegraphics[width=1.0\linewidth]{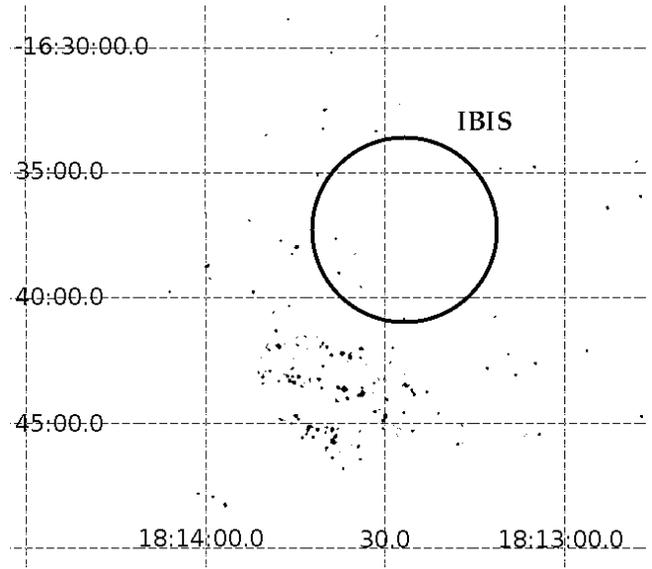}
\caption{XRT 0.3--10 keV image of the region surrounding IGR J18134--1636.
No source is detected by XRT within the IBIS error circle.}
\label{fig16}
\end{figure}

No XRT detection is found at the \emph{Chandra} position nor we detect other sources
within the IBIS error circle (see Figure~\ref{fig16}); thus, we can only infer an  
upper limit to the unabsorbed 0.3--10 keV flux ($\sim$$7\times 10^{-13}$ erg 
cm$^{-2}$ s$^{-1}$) for the entire region, which is lower than the 
\emph{Chandra} flux for CXOO J181328.0--163548, calling for some variability in the X-ray flux of the only 
proposed counterpart suggested so far.

\subsection*{\bf IGR J18175--1530} 
\begin{figure}
\includegraphics[width=1.0\linewidth]{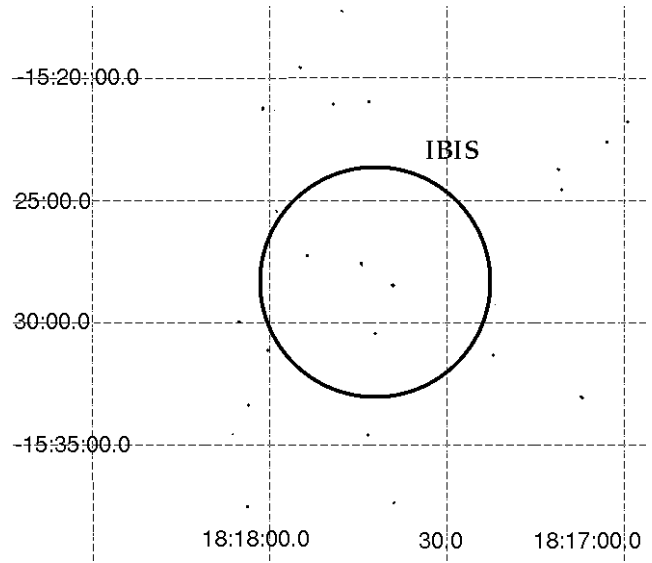}
\caption{XRT 0.3--10 keV image of the region surrounding IGR J18175--1530.
Within the IBIS error circle, no X-ray source is detected by XRT.} 
\label{fig17}
\end{figure}

IGR J18175--1530 was first reported as an IBIS source by Paizis et al.(2007) and
then detected by \emph{RXTE}/PCA during scans of the region (Markwardt et al. 2007).
Indeed, the source is listed in the fourth IBIS catalogue as a transient discovered through
the bursticity analysis. 
Unfortunately, no source is detected in the XRT images of the two observations
available for this region (see Figure~\ref{fig17}), so that again we can only provide an upper limit 
to the 2--10 keV flux ($\sim$$4\times 10^{-13}$ erg cm$^{-2}$ s$^{-1}$) of the X-ray counterpart of 
IGR J18175--1530. Given the extreme X-ray behaviour observed, also in this case we anticipate that it will 
be quite difficult to restrict the source positional uncertainty and provide a unique optical counterpart.

\subsection*{\bf IGR J20569+4940} 
\begin{figure}
\includegraphics[width=1.0\linewidth]{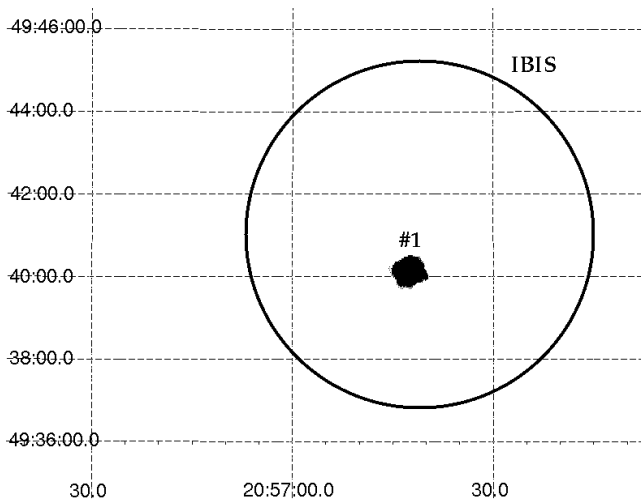}
\caption{XRT 0.3--10 keV image of the region surrounding IGR J20569+4940.
Only one source (\#1) is revealed by XRT within the IBIS error circle.}
\label{fig18}
\end{figure}

This source, also named 3A 2056+493, was first reported by Krivonos et al. (2007) in their all-sky hard 
X-ray survey, but left unclassified.
Within the IBIS circle there is one X-ray source detected at 40$\sigma$ confidence 
level in the longest ($\sim$8.5 ks) XRT pointing (see Figure~\ref{fig18}). It has a counterpart in 
a 2MASS object located at 
RA(J2000) = $20^{\rm h}56^{\rm m}42^{\rm s}.72$, Dec(J2000) = $+49^\circ40^{\prime}06^{\prime \prime}.9$
with magnitudes $J = 13.686\pm0.000$, $H = 14.299\pm0.100$ and $K =13.735\pm0.084$. No optical counterpart  
in the USNO--B1.0 catalogue is associated with this infrared source.
The XRT position is compatible with the ROSAT Bright Survey object 
1RXS J205644.3+494011 and with the \emph{XMM-Newton} Slew Survey source XMMSL1 J205642.7+494004.
\emph{XMM-Newton} detected the source on two occasions reporting a different 0.2--12 keV flux of 
($0.83\pm0.25$) and ($1.85\pm0.31$) $\times10^{-11}$ erg cm$^{-2}$ s$^{-1}$, respectively, on a time-scale 
of hours.
The XRT source also coincides with a relatively powerful radio source (NVSS J205642+494005) having a 
20 cm flux of $167.3\pm5.0$ mJy and detections in various radio catalogues.
Because it is compact and unresolved in radio and has a 2.8 to 11 cm 
flat spectral index of --0.4 (Reich et al. 2000), it is likely a radio loud object.
Indeed, it was selected by Paredes, Ribo \& Martin (2002) as a possible 
microquasar candidate given its location close to the Galactic plane, but a blazar classification cannot 
be excluded at this stage.

The X-ray spectrum is well modelled (see Table~\ref{tab3}) with a slightly absorbed ($N_{\rm H}$ 
$\sim$$5 \times 10^{21}$ cm$^{-2}$) power law with a photon index $\Gamma$ $\sim$2.3 and a 2--10 keV flux 
of $\sim$$1 \times10^{-11}$ erg cm$^{-2}$ s$^{-1}$.

Optical follow-up observations are required to define if this is a Galactic or extragalactic jet 
source.

\subsection*{\bf 1RXS J213944.3+595016} 
\begin{figure}
\includegraphics[width=1.0\linewidth]{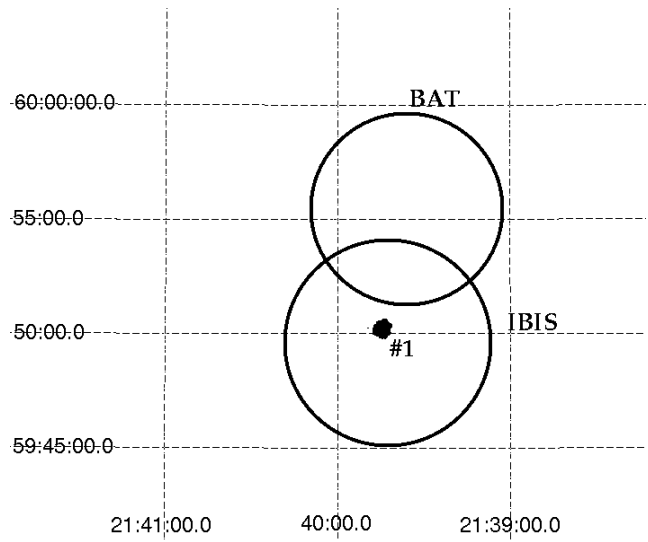}
\caption{XRT 0.3--10 keV image of the region surrounding 1RXS J213944.3+595016.
The only X-ray object (source \#1) detected by XRT is well located within the IBIS unceratinty (larger
black circle) and just outside the BAT error circle (smaller black circle).}
\label{fig19}
\end{figure}

This source is also reported in the \emph{Swift}/BAT survey of Cusumano et al. (2009), where it is 
associated to a ROSAT Bright Survey object detected within the IBIS 
error circle and just outside the BAT positional uncertainty; it is located with an accuracy of 7 arcsec.
The same source is also the only detection found in the XRT image (see Figure~\ref{fig19}) and it is 
clearly the counterpart of the IBIS and possibly BAT object. It is quite bright in X-rays 
($\sim$16$\sigma$ confidence 
level) and localised with a better accuracy than the one provided by ROSAT ($\sim$4 arcsec). Within this 
uncertainty lies a USNO--B1.0 object located at RA(J2000) = $21^{\rm h}39^{\rm m}45^{\rm s}.20$, Dec(J2000) 
= $+59^\circ50^{\prime}14^{\prime \prime}.7$ (with magnitude $R$ $\sim$18.7), which is also listed in the 
2MASS survey with magnitudes $J = 15.165\pm0.056$, $H = 14.067\pm0.062$ and $K = 12.889\pm0.038$.

The X-ray spectrum, fitted with a slightly absorbed ($N_{\rm H}$ $\sim$$3.5 \times 10^{21}$ 
cm$^{-2}$) power law with a photon index $\Gamma$ $\sim$2 and a 2--10 keV flux of
$\sim$$6 \times10^{-12}$ erg cm$^{-2}$ s$^{-1}$ (see Table~\ref{tab3}), is similar to those typically 
observed in AGNs. Indeed, recent optical follow-up measurements confirm the AGN classification
(Masetti et  al. 2009).

\subsection*{\bf IGR J22234--4116} 
\begin{figure}
\includegraphics[width=1.0\linewidth]{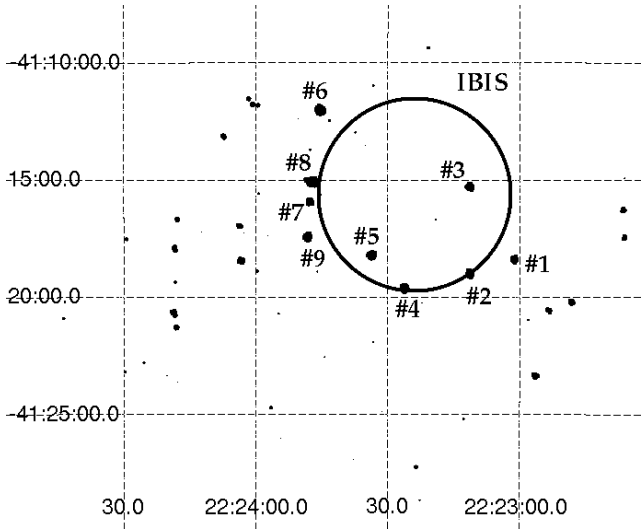}
\caption{XRT 0.3--10 keV image of the region surrounding IGR J22234--4116.
Within the 90\% IBIS error circle, XRT detects five X-ray sources of which,
two (\#3 and \#5) well inside it, and three (\#2, \#4 and \#8) at its border. The remaining four sources 
(\#1, \#6, \#7 and \#9) are instead located within the 99\% IBIS uncertainty.}
\label{fig20}
\end{figure}

Although this source is reported as persistent both in the third and fourth IBIS 
catalogues (Bird et al. 2007; 2009), the XRT data do not show the clear presence 
of a bright source within the IBIS error circle (see Figure~\ref{fig20}). Instead the analysis of 
the XRT image provide evidence 
for 9 detections of which six are located within the 90\% and three within the 99\% IBIS error circle 
(see Table~\ref{tab2}). All these sources disappear above 3 keV indicative of a soft spectral shape and 
only a few objects have possible counterparts at other wavelengths. NED also reports the presence in the 
region of a loose group of galaxies but none of its members is detected by XRT (Tucker et al 2000).

Source \#2 has a counterpart in a USNO B--1.0 catalogue located at RA(J2000) =
$22^{\rm h}23^{\rm m}11^{\rm s}.33$, Dec(J2000) = $-41^\circ19^{\prime}01^{\prime \prime}.0$
(with magnitude $R$ $\sim$18.9--19.1) but it is not listed in the 2MASS survey.
Its XRT position is also compatible with a ROSAT Faint object (1RXS J222313.2--411923).
Within the XRT uncertainty of source \#6, we find a radio source belonging to the SUMSS catalogue
(SUMSS J222344--411150) having a 36 cm flux of $20.4\pm1.8$ mJy.
In the case of source \#9, the XRT position is consistent with those of a USNO--B1.0 object 
located at RA(J2000) = $22^{\rm h}23^{\rm m}48^{\rm s}.26$, Dec(J2000) = $-41^\circ17^{\prime}27^{\prime 
\prime}.1$ (with magnitude $R$ $\sim$18) also listed in the 2MASS survey with magnitudes $J = 
15.561\pm0.049$, $H = 15.132\pm0.070$ and $K = 14.965\pm0.106$. 

All these objects are too faint in X-rays to perform any spectral analysis, 
allowing only a flux estimate in the 2--10 keV energy band in the range $\sim$($0.6-3)\times10^{-13}$ erg 
cm$^{-2}$ s$^{-1}$.

Although we are not able to pinpoint amongst all these sources the true counterpart of the IBIS source,
we note that IGR J22234--4116 is most likely an extragalactic object on the basis of its high Galactic
latitude ($b \sim -56^{\circ}$) as we rarely find a cluster of galaxy as an IBIS source.

\begin{table}
\caption{Summary of the proposed counterparts.}
\label{tab4} 
\begin{flushleft}
\begin{tabular}{lc}
\hline
\hline
Source & Type \\
\hline
\hline
IGR J00465--4005             &    AGN, absorbed       \\ 
LEDA 96373                   &    AGN, Seyfert 2          \\
IGR J07506--1547             &    unidentified       \\
1RXS J080114.6--462324       &    Galactic source    \\ 
IGR J08262+4051              &    AGN candidate      \\
IGR J10447--6027             &    unidentified       \\
MCG+04--26--006              &    AGN, LINER              \\
PKS 1143--696                &    AGN candidate, QSO?      \\
IGR J12123--5802             &    unidentified       \\
IGR J1248.2--5828            &    AGN, absorbed ?       \\
NGC 4748                     &    AGN, NLS1         \\
IGR J13107--5626             &    AGN, absorbed ?     \\
IGR J14080--3023             &    AGN, Seyfert 1.5     \\
IGR J17008--6425             &    unidentified       \\
IGR J17331--2406             &    unidentified       \\
IGR J18134--1636             &    unidentified       \\
IGR J18175--1530             &    unidentified       \\
IGR J20569+4940              &    Blazar or microQSO      \\
1RXS J213944.3+595016        &    AGN, unabsorbed       \\
IGR J22234--4116             &    AGN candidate      \\
\hline
\hline
\end{tabular}
\end{flushleft}
\end{table}

\section{Conclusions}

The use of the XRT data archive has allowed us to find the optical counterpart 
and/or provide X-ray information for 20 \emph{INTEGRAL} sources in the fourth IBIS catalogue. 
The result of our work is summarised in Table~\ref{tab4}.
Interestingly, the majority of the sources discussed in this paper appear to be AGN
even if their location is sometime close to the Galactic plane. Four objects
(IGR J00465--4005, LEDA 96373, IGR J1248.2--5828 and IGR J13107--5626) are confirmed or likely 
absorbed active galaxies, while two (IGR J14080--3023 and 1RXS J213944.3+595016) are unabsorbed AGN. 
Three objects are more peculiar extragalactic objects, NGC 4728 being a Narrow Line Seyfert galaxy, 
MCG+04--26--006 a type 2 LINER and PKS 1143--693 probably a QSO; finally, two objects (IGR J08262+4051, 
and IGR J22234--4116) are candidate AGN, which
require further optical spectroscopic follow-up observations to be classified.
Only in the case of 1RXS J080114.6--462324 we are confident that the source is a Galactic object.
In three cases (IGR J10447--6027, IGR J12123--5802 and IGR J20569+4940) one X-ray counterpart was 
pinpointed, although its nature could not be assessed despite spectral and sometimes variability 
information have been obtained. Clearly, these objects need to be optically observed and classified
before their nature is firmly assessed. Finally, in five cases no obvious X-ray counterpart 
(IGR J07506--1547 and IGR J17008--6425) or even no detection (IGR J17331--2406, IGR J18134--1636 and
IGR J18175--1530) was found; with the exception of IGR J18134--1636, 
all these sources are variable in the IBIS band and hence difficult to catch even 
at X-ray energies. This poses a problem in the search for their optical counterpart as ``standard'' 
follow-up observations will be useless unless a well defined monitoring strategy 
can be conceived and considerable amount of observing time obtained.

For most sources in the sample, X-ray spectral properties have been obtained 
for the first time, while flux estimates (either in case of positive detection or as upper limits)
have been reported for all objects/regions analysed.
Where possible, flux variability between observations have been obtained and discussed.
As a final remark, we note that the results of this work confirm the key role played by follow-up 
observations with current X-ray telescopes and how it is important to encourage multiwaveband studies, in 
particular optical spectroscopy, which will help to definitely assess the nature of as yet unidentified 
high energy sources.

\section*{Acknowledgments}
This research has made use of data obtained from the SIMBAD database operated at CDS, Strasbourg, France; 
the High Energy Astrophysics Science Archive Research Center (HEASARC), provided by NASA's Goddard Space 
Flight Center NASA/IPAC Extragalactic Database (NED). We thank the anonymous referee for useful 
remarks which helped us to improve the quality of this paper.
We also acknowledge the use of public data from the \emph{Swift} data archive.
The authors acknowledge the ASI financial support via ASI--INAF grant I/008/07/0.

\end{document}